\documentclass[a4paper,aps,amsfonts,amsmath,nofootinbib,showpacs,byrevtex,preprintnumbers,showkeys,prd,10pt%
,notitlepage%
,reprint%
]{revtex4-1}

\usepackage[dvips]{hyperref}
\usepackage[dvips]{graphicx}
\usepackage{bm}

\usepackage[caption=false]{subfig}

\DeclareMathOperator{\Det}{Det}
\DeclareMathOperator{\Tr}{Tr}

\renewcommand{\i}{\mathrm{i}}
\newcommand{\e}{\mathrm{e}}

\newcommand{\cN}{{\mathcal{N}}}
\newcommand{\cD}{{\mathcal{D}}}
\newcommand{\cF}{{\mathcal{F}}}
\newcommand{\cS}{{\mathcal{S}}}
\newcommand{\cO}{{\mathcal{O}}}

\newcommand{\be}{\begin{equation}}
\newcommand{\ee}{\end{equation}}

\newcommand{\bra}[1]{\langle #1\rvert}
\newcommand{\ket}[1]{\lvert#1\rangle}
\newcommand{\braket}[2]{\langle #1 \vert #2 \rangle}
\newcommand*{\vev}[1]{\left< #1 \right>}
\def\abs#1{\lvert#1\rvert}

\providecommand*{\coloneq}{\mathrel{\mathop:}=}
\providecommand*{\eqcolon}{=\mathrel{\mathop:}}

\newcommand{\rk}{\right)}
\newcommand{\lk}{\left(}

\renewcommand*{\vec}[1]{\bm{\mathrm{#1}}}

\newcommand{\vA}{\vec{A}}
\newcommand{\vx}{\vec{x}}
\newcommand{\vy}{\vec{y}}

\newcommand{\vk}{\vec{k}}
\newcommand{\vp}{\vec{p}}
\newcommand{\vq}{\vec{q}}

\newcommand{\kbol}{k^{}_{\mathrm{B}}}

\newcommand*{\Eqref}[1]{Eq.~\eqref{#1}}

\renewcommand*{\d}[1][]{\mathop{\mathrm{d}^{#1}}\mkern-4mu}
\newcommand*{\dbar}[1][]{\mathop{\mathrm{d}\mkern-7mu\mathchar'26\mkern-1mu^{#1}}\mkern-4mu}

\begin{document}

\title{The deconfinement phase transition in the Hamiltonian approach to Yang--Mills theory in Coulomb gauge}

\author{Jan Heffner}
\author{Hugo Reinhardt}
\author{Davide R.~Campagnari}
\affiliation{Institut f\"ur Theoretische Physik, Universit\"at T\"ubingen,
Auf der Morgenstelle 14, 72076 T\"ubingen, Germany}
\date{29 February 2012}
\pacs{%
11.10.Wx, 
11.15.-q, 
12.38.Aw, 
12.38.Lg 
}
\keywords{Hamiltonian approach; Coulomb gauge; Deconfinement; Yang-Mills Theory}

\begin{abstract}
The deconfinement phase transition of SU(2) Yang--Mills theory is investigated
in the Hamiltonian approach in Coulomb gauge assuming a quasi-particle picture for the
grand canonical gluon ensemble. The thermal equilibrium state is found by minimizing the
free energy with respect to the quasi-gluon energy. Above the deconfinement phase
transition the ghost form factor remains infrared divergent but its infrared exponent is approximately
halved, while the gluon energy, being infrared divergent in the confined phase,
becomes infrared finite in the deconfined phase. For the effective gluon mass we
find a critical exponent of $0.37$. Using the lattice results for the gluon propagator
to fix the scale, the deconfinement transition temperature is obtained in the range of $275$ to $290$ MeV. 
\end{abstract}

\maketitle

\section{Introduction}\label{section1}

Understanding the phase diagram of Quantum Chromodynamics (QCD) is one of the major
challenges of particle physics. Running and upcoming, respectively, high-energy heavy-ion
experiments at LHC, RHIC, SPS and FAIR, NICA and J-PARC call for a deeper understanding
of hadronic matter under extreme conditions. The central issues are the equation of state
of QCD and the nature of the phase transition from the confined hadronic phase with
chiral symmetry spontaneously broken to the deconfined quark gluon plasma with chiral
symmetry restored. The deconfinement phase transition is expected to be driven by the
gluon dynamics, while the chiral phase transition, obviously, is due to strong
interaction of the quarks, which, of course, is also mediated by the gluons. It is
therefore by far non-trivial that both phase transitions occur at roughly the same
temperature as observed on the lattice \cite{Karsch:2001cy}.

In quenched QCD the deconfinement phase transition is related to the center of the gauge
group. Center symmetry is realized in the low-temperature confined phase and spontaneously
broken in the high-temperature deconfined phase \cite{Svetitsky:1982gs}. When quarks are
included center symmetry is explicitly broken and the deconfinement transition is
expected to become a cross-over. The features of the chiral phase transition dominantly
depend on the quark masses, which explicitly break chiral symmetry, but also on the
strength of the chiral anomaly. Furthermore, lattice calculations also show that confinement
is generated exclusively by the low-energy gluonic modes while also medium-energy gluon
modes contribute to the order parameter of spontaneous breaking of chiral symmetry, the
quark condensate \cite{Yamamoto:2009de}.

Meanwhile by various theoretical studies evidence has been accumulated that the phase structure
of hadronic matter is much richer than originally thought \cite{BraunMunzinger:2009zz}. As the chemical potential or baryon
density is increased one expects, on the basis of large-$N_c$ arguments, the existence of a ``quarkyonic phase''
where quarks and gluons are still confined but chiral symmetry is restored \cite{McLerran:2007qj}. At even larger
chemical potential the transition to color superconducting quark matter occurs with color-flavor
locking \cite{Alford:2007xm}. There are also speculations on the existence of a chiral critical endpoint for finite baryon
density \cite{Fodor:2001pe}.

Clearly the understanding of the phase structure and, in particular, the deconfinement phase transition 
requires, like the understanding of the confined phase itself, non-perturbative methods.
A rigorous non-perturbative treatment of QCD is achieved on the lattice. The lattice method has been 
quite successful for quenched QCD but becomes extremely expensive when dynamical quarks are 
included and fails at large baryon densities due to the notorious fermion sign problem at non-vanishing
chemical potential. Therefore, alternative non-perturbative methods which rely on the continuum
formulation of QCD and thus do not suffer from the problem connected with the lattice treatment
of fermions are very desirable. In recent years substantial progress has been made in first principle
continuum QCD calculations, which rely on functional methods and do not suffer from the fermion problems
of the lattice. Among these methods are Dyson--Schwinger equations (DSEs) in Landau \cite{Fischer:2006ub}
and Coulomb gauge \cite{Watson:2006yq}, functional renormalization group
(FRG) flow equations in Landau gauge \cite{Pawlowski:2005xe} and Hamiltonian Coulomb gauge \cite{Leder:2010ji}, and
variational approaches to the Hamilton formulation 
of Yang--Mills theory in Coulomb gauge \cite{Schutte:1985sd,*Szczepaniak:2001rg,Feuchter:2004mk}.
These various continuum methods have all their own advantages and drawbacks, 
and by combining them one expects to gain new insights into the non-perturbative regime of the theory. 
So far the deconfinement phase transition has been studied mainly in the FRG approach
\cite{Pawlowski:2010ht,Braun:2007bx} or in the DSE approach \cite{Fischer:2009wc}.
The deconfinement temperature was extracted either directly from the Polyakov loop 
\cite{Braun:2007bx}, which (in the absence of quarks) is the order parameter of
confinement, or from the dual condensate \cite{Fischer:2009wc}, which can be related to
the Polyakov loop \cite{Gattringer:2006ci}.

In this paper we study this transition in a variational approach to Hamiltonian Yang--Mills theory extending previous work 
\cite{Feuchter:2004mk} to finite temperature.

In the variational approach at zero temperature the energy density is minimized using Gaussian
type ans\"atze for the Yang--Mills vacuum wave functional \cite{Schutte:1985sd,Szczepaniak:2001rg,Feuchter:2004mk}. So far, this approach has been mainly applied
to study the infrared sector of Yang--Mills theory \cite{Schleifenbaum:2006bq,Epple:2006hv}, but was recently extended to full QCD \cite{Pak:2011wu}. In 
the present paper we extend this approach to finite temperatures and study the deconfinement phase
transition. For this purpose we will minimize the free energy after making an appropriate ansatz for the
density matrix. Some initial investigations in this direction 
were undertaken in Ref.~\cite{Reinhardt:2011hq}, where, for simplicity, only the so-called subcritical solutions \cite{Epple:2007ut} were
considered, which do not produce confinement in the sense that they yield an infrared (IR) finite gluon energy and
not a linearly rising (static) quark potential. As a consequence the deconfinement phase transition 
could not be studied. 
Furthermore, in Ref.~\cite{Reinhardt:2011hq} the projection of the grand canonical gluon
ensemble onto zero color was considered and it was found that the effect of color projection is
negligible. We will therefore ignore color projection in the present paper.

The organization of the paper is as follows: In Section~\ref{section2} we briefly summarize
the basic ingredients of the Hamiltonian approach to Yang--Mills theory in Coulomb gauge,
which are needed for its extension to finite temperatures. In Section~\ref{section3b} we
introduce the grand canonical ensemble of Yang--Mills theory. We define here
the complete basis of the gluon Fock space as well as the density matrix following
Ref.~\cite{Reinhardt:2011hq}. In Section~\ref{section4}
we present the DSEs for the ghost and Coulomb propagator. In Section~\ref{section5} we calculate
the partition function, the entropy and the free energy. The variation of the free energy is
carried out in Section~\ref{section6} and the renormalization of the resulting DSEs is
given in Section~\ref{section7}. 
In Section~\ref{section8} we present the results of an infrared analysis of the coupled DSEs and discuss the importance of the Coulomb
term of the Yang--Mills Hamiltonian in Section~\ref{section9}. 
Finally, in Section~\ref{section10} we present the numerical solution of these 
DSEs and determine, in particular, the deconfinement transition temperature. 
A short summary and our conclusions are given in Section~\ref{section11}. The details of the IR analysis are presented in 
the Appendix. 


\section{Hamiltonian approach to Yang--Mills theory in Coulomb gauge}\label{section2}

Below we briefly summarize the basic ingredients of the Hamiltonian approach to Yang--Mills
theory in Coulomb gauge \cite{Feuchter:2004mk} needed for its extension to finite
temperatures \cite{Reinhardt:2011hq}.

To simplify the bookkeeping we will use the compact notation  $A^{a_1}_{k_1} (\vx_1)  =  A (1)$
for colored Lorentz vectors like the gauge potential and an analogous notation for colored
Lorentz scalars like the ghost $C^{a_1} (\vx_1) = C (1)$. A repeated label means summation
over the discrete color (and Lorentz) indices along with integration over the $d$
spatial coordinates
\be
A \cdot B = A (1) B (1) = \int \d[d]x \sum_{i=1}^{d} \sum_{a=1}^{N_c^2-1} A^a_i (\vx) B^a_i (\vx) .
\ee
We define in coordinate space
\be\label{279-xx}
\delta(1, 2) = \delta^{a_1 a_2} t_{i_1 i_2} (\vx_1) \, \delta (\vx_1-\vx_2) , 
\ee
where
\be
\label{284-xx}
t_{ij} (\vx) = \delta_{ij} - \frac{\partial_i \partial_j}{\partial^2}
\ee
is the transverse projector. Furthermore, indices will be suppressed when they can be
easily restored from the context.

After resolving Gauss's law in Coulomb gauge the Yang--Mills Hamiltonian \cite{Christ:1980ku}
reads in our notation
\be\label{398-G1}
\begin{split}
H ={}& \frac{1}{2} \bigl[ J^{-1}_A \Pi(1) J_A \Pi(1) + B(1) \, B(1) \bigr]  \\
&+ \frac{g^2}{2} J^{-1}_A \rho(1) J_A \, F_A(1, 2) \rho(2) \\
\equiv{}& H_K + H_B + H_\mathrm{C} , 
\end{split}
\ee
where $\Pi(1) = -\i \delta/\delta A(1)$ is the canonical momentum (electric field)
operator, $B(1)$ is the non-abelian magnetic field
\be\label{magfield}
\vec{B}^a = \vec{\nabla} \bm{\times} \vA^a + \frac{g}{2} \: f^{abc} \vA^b \bm{\times} \vA^c  \qquad (d=3),
\ee
$g$ is the coupling constant, and
\be\label{404}
\begin{split}
J_A &= \Det G_A^{-1}, \\
G^{-1}_A(1,2) &= \bigl( -\delta^{a_1a_2} \, {\partial^2_{\vx_1}} - g \, \hat{A}^{a_1a_2}_i(\vx_1) {\partial_i^{\vx_1}} \bigr) \delta(\vx_1-\vx_2)
\end{split}
\ee
is the Faddeev--Popov determinant with $\hat{\vA}{}^{ab} = f^{acb} \vA^c$ being the
gauge field in the adjoint representation of
the gauge group SU($N_c$) with structure constants $f^{abc}$. Furthermore,
\be\label{409-G3}
\begin{split}
\rho (1) &\equiv \rho^{a_1}(\vx_1) = R(1;2,3) A(2) \Pi(3), \\
R(1;2,3) &= f^{a_1a_2a_3} \, \delta_{i_2 i_3} \, \delta(\vx_1-\vx_2) \, \delta(\vx_1-\vx_3)
\end{split}
\ee
is the color charge density of the gluons and
\be\label{414-G4}
F_A (1, 2) \equiv F^{a_1 a_2}_A (\vx_1, \vx_2) = G_A(1,3) \, G_0^{-1}(3,4) \, G_A(4,2)
\ee
is the so-called Coulomb kernel, with $G_0^{-1}$ being the bare inverse ghost operator, obtained by
setting $A=0$ in Eq.~\eqref{404}. In the presence of matter fields with color charge
density $\rho_m (1)$, the gluon charge $\rho(1)$ in the Coulomb term $H_\mathrm{C}$
[\Eqref{398-G1}] is replaced by the total charge $\rho (1) + \rho_m (1)$ and the vacuum
expectation value of $F_A(1,2)$ acquires the meaning of the static non-abelian Coulomb
potential.

The gauge fixed Hamiltonian Eq.~\eqref{398-G1} is highly non-local due to Coulomb kernel
$F_A(1, 2)$, Eq.~\eqref{414-G4}, and due to the presence of the Faddeev--Popov determinant
$J_A$, Eq.~\eqref{404}. In addition, the latter also occurs in the functional integration
measure of matrix elements of operators $\cO$ between Coulomb gauge wave functionals
\be\label{419-G5}
\langle \psi_1 | \cO | \psi_2 \rangle = \int D A \, J_A \, \psi^*_1[\vec{A}] \, \cO \, \psi_2 [\vec{A}] ,
\ee
where the integration runs over transverse field configurations.
In practical calculations the elimination of unphysical degrees of freedom via gauge
fixing is usually beneficial (and sometimes unavoidable), in spite of the increased
complexity of the gauge-fixed Hamiltonian Eq.~(\ref{398-G1}). The non-trivial
Faddeev--Popov determinant reflects the intrinsically non-linear structure of the space
of gauge orbits and dominates the IR behavior of the theory. Once Coulomb gauge is
implemented, any functional of the (transverse) gauge field is a physical state. 


\section{The grand canonical ensemble of Yang--Mills theory}\label{section3b}

We are interested in the behavior of Yang--Mills theory at finite temperatures. For this
purpose we consider the grand canonical ensemble of Yang--Mills theory, which is defined
in the Hamiltonian approach by the density matrix 
\be
\label{7-x}
\cD = \exp (- \beta H)
\ee
with $\beta = 1/\kbol T$ being the inverse temperature measured in units of the Boltzmann
constant $\kbol$. Formally, $\cD$ [Eq.~(\ref{7-x})] looks like the canonical ensemble
since the chemical potential of the gluon vanishes. However, the thermal expectation values 
\be
\label{6-x}
\langle\!\langle \cO \rangle\!\rangle = \frac{\Tr\cD \cO }{\Tr\cD} = \frac{\sum_k \bra{k} \cD \cO \ket{k}}{\sum_k \bra{k} \cD \ket{k}}
\ee
have to be taken over the whole Fock space, i.e.\ over states $\{\ket{k}\}$ with an arbitrary number
of gluons. Since $H$ is non-linear and non-local it is clear from the very beginning that
we have to resort to approximations. We will use an analogous approximation scheme as in
the zero-temperature case, see Ref.~\cite{Feuchter:2004mk}. 

The trace in Eq.~(\ref{6-x}) can, in principle, be calculated in any complete basis $\{\ket{k}\}$. Since
we are mainly interested in $\langle\!\langle H \rangle\!\rangle$ we will choose a basis
which is adapted to the structure of the Yang--Mills Hamiltonian. Following Refs.~\cite{Feuchter:2004mk}
and \cite{Reinhardt:2011hq} we choose the basis of the gluonic Fock space of the form
\be\label{8-x}
\braket{A}{k} = J^{-1/2}_A \braket{A}{\tilde{k}} ,
\ee
where $\ket{\tilde{k} }$ denotes a complete set of states of the Yang--Mills Fock space
to be specified later. This ansatz removes the Faddeev--Popov determinant from the integration measure in Eq.~(\ref{419-G5}).  
In this basis the thermal expectation value Eq.~(\ref{6-x}) reads
\be\label{9-x}
\langle\!\langle \cO \rangle\!\rangle =
\frac{\sum_k \bra{\tilde{k}} \tilde{\cD} \tilde{\cO} \ket{\tilde{k}} }{\sum_k \bra{\tilde{k}} \tilde{\cD} \ket{\tilde{k}}} \eqcolon
\langle \tilde{\cO} \rangle,
\ee
where we have introduced the abbreviation
\be\label{10-x}
\tilde{\cO} = J^{1/2}_A \cO J^{- 1/2}_A ,
\ee
which yields for the density matrix Eq.~(\ref{7-x})
\be\label{337}
\tilde{\cD} = J^{1/2}_A \cD J^{- 1/2}_A = \e^{- \beta \tilde{H}} .
\ee
The transformed Yang--Mills Hamiltonian $\tilde{H}$ defined by Eq.~(\ref{10-x}) is
obtained by replacing in the Yang--Mills Hamiltonian Eq.~(\ref{398-G1}) the momentum
operators $J^{-1}_A \Pi J_A \Pi$ by $\tilde{\Pi}^\dagger \tilde{\Pi}$, where\footnote{Note
that since the color density of the gluon field $\rho$ (\ref{409-G3}) is linear
in the momentum operator $\Pi$, the Coulomb Hamiltonian
$H_\mathrm{C}$ (\ref{398-G1}) has, concerning the occurrence of $\Pi$ and $J_A$, the same structure as the kinetic 
term $\sim J^{- 1}_A \Pi J_A \Pi$. This is not surprising since the Coulomb term is the longitudinal part of the 
kinetic energy
$\sim J^{- 1}_A \Pi^{\parallel} J_A \Pi^{\parallel}$ with $\vec{\Pi}^{\parallel}$ being the longitudinal momentum operator determined by 
the resolution of Gauss's law.}
\be
\label{11-1}
\tilde{\Pi} = J^{1/2}_A \Pi J^{-1/2}_A = \Pi + \frac{\i}{2} \frac{\delta \ln J_A}{\delta A}  \, .
\ee
To work out the thermal averages it is convenient to go to momentum space
\be\label{28-x}
\begin{split}
A (\vx) &= \int \dbar{k} \: \e^{\i \vk\cdot\vx} \, A(\vk) , \\
\Pi (\vx) &= \int \dbar{k} \: \e^{-\i \vk\cdot\vx} \, \Pi(\vk) ,
\end{split}
\ee
($\dbar{k} \equiv \d[d]k/(2 \pi)^d$) where the Fourier components satisfy the commutation relation
\be
\label{29-x}
\bigl[ A(1) , \Pi(2) \bigr] = \i \, \delta(1, 2)  .
\ee
In the standard fashion we express the Fourier components in terms of creation and annihilation operators
\begin{subequations}\label{32-33-x}
\begin{align}
\label{32-x}
A_i^a(\vk) & = \frac{1}{\sqrt{2 \omega (\vk)}} \lk a_i^a(\vk)  + a_i^{a\dagger} (-\vk) \rk , \\
\label{33-x}
\Pi_i^a(\vk) & = \i \sqrt{\frac{\omega (\vk)}{2}} \lk a_i^{a\dagger}(\vk) - a_i^a(-\vk) \rk ,
\end{align}
\end{subequations}
which satisfy the usual commutation relations
\be
\label{34-x}
\bigl[a_i^a(\vk) , a_j^{b\dagger} (\vq) \bigr] = \delta^{ab} t_{ij}(\vk) (2\pi)^d \delta(\vk-\vq), 
\ee
where $t_{ij} (\vk) = \delta_{ij} - \hat{k}_j \hat{k}_j$, $\hat{\vk} = \vk/\abs{\vk}$ 
is the transverse projector Eq.~\eqref{284-xx} in momentum space.
Here $\omega (\vk)$ is, so far, an arbitrary (positive definite) function of momenta. It 
enters the vacuum state $\ket{\tilde{0}}$ defined by
\be
\label{35-x}
a_i^a(\vk) \ket{\tilde{0}} = 0 ,
\ee
which has the ``coordinate'' representation
\be
\label{36-x}
\braket{A}{\tilde{0}} = \cN \exp \left[ - \tfrac12 A(1) \omega(1, 2) A(2) \right] ,
\ee
where $\cN$ is a normalization constant. The states
\be
\label{37-x}
\ket{\tilde{0}}, \quad
a_i^{a\dagger}(\vk)\ket{\tilde{0}}, \quad
a_i^{a\dagger}(\vk) \, a_j^{b\dagger}(\vq) \ket{\tilde{0}}, \quad\dots
\ee
form a complete basis in the gluonic Fock space for any positive definite function
$\omega(\vk)$. Below we will use this basis to evaluate the thermal expectation values
Eq.~(\ref{9-x}).

Because of the non-local structure of the Yang--Mills Hamiltonian [Eq.~(\ref{398-G1})],
the density matrix Eq.~(\ref{337}) can only be treated in an approximate fashion. To make
the actual calculation feasible we follow Ref.~\cite{Reinhardt:2011hq} and replace the full
Yang--Mills Hamiltonian $\tilde{H}$ in the density operator $\tilde{\cD}$ [Eq.~(\ref{337})]
by a single-particle operator
\be\label{15}
h = \int \dbar{k} \: a^{a\dagger}_i (\vk) \, \Omega^{ab}(\vk) \, a^b_i (\vk), 
\ee
where $\Omega^{ab}(\vk)$ will be later determined by minimizing the energy density,
yielding
\be\label{271}
\tilde{\cD} = \exp (- \beta h) .
\ee
By global color and rotational invariance, $\Omega^{ab} (\vk)$ is color-diagonal and
independent of color, $\Omega^{ab}(\vk) = \delta^{ab} \Omega(\vk)$, and furthermore
depends only on $\abs{\vk}$. The same is true for the kernel $\omega$ in the vacuum wave
functional Eq.~(\ref{36-x}). Since the transformed density matrix $\tilde{\cD}$
[Eq.~(\ref{271})] is the exponential of a single-particle operator Wick's theorem applies
to the thermal averages Eq.~(\ref{9-x}), whose temperature dependence is exclusively given
by the finite-temperature Bose occupation numbers $n(\vk)$, defined (in momentum space) by
\be\label{482-X1}
\begin{split}
\langle a_i^{a\dagger}(\vk) \, a_j^b(\vq) \rangle &= \delta^{ab} t_{ij}(\vk) (2\pi)^d \delta(\vk-\vq) \, n(\vk) , \\
\qquad n (\vk) &= \bigl[ \exp\bigl(\beta\Omega(\vk)\bigr) - 1 \bigr]^{- 1} .
\end{split}
\ee
From Eqs.~(\ref{32-x}) and (\ref{33-x}) one finds for the gluon propagator
\be
\label{488-x2}
D(1, 2) \coloneq \langle A (1) A ( 2) \rangle , \qquad
D(\vk) =  \frac{1+ 2 n (\vk)}{2 \omega (\vk)} ,
\ee
for the momentum propagator
\be
\label{494-x3}
K (1, 2) \coloneq \langle \Pi (1) \Pi (2) \rangle , \qquad
K(\vk) = \frac{1 + 2n (\vk)}{2} \: \omega (\vk) ,
\ee
and for the ``mixed'' propagator
\be
\label{500-x4}
\langle A (1) \Pi (2) \rangle = \frac{\i}{2} \: \delta (1, 2) ,
\ee
the latter being independent of the temperature.


\section{Ghost and Coulomb propagator}\label{section4}

We will resort to the same approximation used in Ref.~\cite{Feuchter:2004mk} in the $T = 0$ case, i.e.\
calculating the energy up to two loops. To this order the following representation of the
Faddeev--Popov determinant holds \cite{Reinhardt:2004mm}
\be\label{430}
J_A = \exp \left[-A (1) \chi (1, 2) A (2) \right] ,
\ee
where
\be
\label{455-38}
\chi (1, 2) = - \frac{1}{2} \vev{ \frac{\delta^2 \ln J_A}{\delta A (1) \delta A (2)} }
\ee
is the so-called curvature, which, in fact, represents the ghost loop (see below). Strictly
speaking this representation was derived only for $T = 0$. However, the proof given in
Ref.~\cite{Reinhardt:2004mm} can be straightforwardly extended to finite temperatures provided the
density matrix $\tilde{\cD}$ is the exponential of a single particle operator, 
so that Wick's theorem holds for the thermal averages $\langle \tilde{\cO} \rangle$ [Eq.~(\ref{9-x})].
This is the case for the density matrix Eq.~(\ref{271}). With the inverse Faddeev--Popov
operator $G_A$ [see Eq.~\eqref{404}] the curvature Eq.~(\ref{455-38}) can be expressed as
\be
\label{291}
\chi (1, 2) = \frac{1}{2} \vev{ \widetilde{\Gamma}_0(1;3,4) \, G_A(3',3) \, \widetilde{\Gamma}_0(2;4',3') \, G_A(4,4') } ,
\ee
where
\be
\label{296}
\widetilde{\Gamma}_0(1;2,3) = \frac{\delta G_A^{- 1}(2,3)}{\delta A (1)}
\ee
is the bare ghost-gluon-vertex. Defining the ghost propagator by
\be
\label{301}
G = \langle G_A \rangle
\ee
and the full ghost-gluon vertex $\widetilde{\Gamma}$ by
\be
\label{306}
\langle G_A \widetilde{\Gamma}_0 G_A \rangle = G \widetilde{\Gamma} G ,
\ee
the ghost loop Eq.~(\ref{291}) becomes
\be
\label{311}
\chi(1, 2) = \frac{1}{2} \widetilde{\Gamma}_0(1;3,4) \, G(3',3) \, \widetilde{\Gamma}(2;4',3') \, G(4,4') .
\ee
Furthermore, the ghost propagator $G$ satisfies the Dyson equation \cite{Feuchter:2004mk}
\be
\label{ha-300}
G = G_0 + G_0 \Sigma G ,
\ee
where
\be
\label{ha-305}
G^{ab}_0(\vx,\vy) = \bigl[ (- \partial^2)^{- 1} \bigr]_{\vx,\vy}^{ab}
\ee
is the free ghost propagator and
\be
\label{ha-310}
\Sigma (1, 2) = \widetilde{\Gamma}_0(3;1,4) \, G (4,4') \, D(3,3') \, \widetilde{\Gamma}(3';4',2)
\ee
is the ghost self-energy. As shown in Landau gauge \cite{Taylor:1971ff} the ghost-gluon vertex is
not renormalized; this applies also in Coulomb gauge, see Ref.~\cite{Schleifenbaum:2006bq}. 
At zero temperature the full ghost-gluon vertex $\widetilde{\Gamma}$ can be, to good
approximation, replaced by the bare one $\widetilde{\Gamma}_0$ \cite{Campagnari:2011bk}. We assume that
this approximation works also at finite temperatures. Assuming a bare ghost-gluon vertex
and expressing the ghost propagator Eq.~(\ref{301}) in momentum space by the ghost form factor
\be
\label{29}
G^{ab}(\vk) = \delta^{ab} \frac{d(\vk)}{g \, \vk^2}
\ee
the curvature Eq.~(\ref{311}) becomes in momentum space
\be
\label{30}
\chi(\vk) = \frac{N_c}{2(d-1)} \int \dbar{q} \bigl[ 1 -  (\hat{\vk} \cdot \hat{\vq})^2 \bigr] 
\frac{d(\vk - \vq) \, d(\vq)}{(\vk - \vq)^2} ,
\ee
and the ghost form factor $d(\vk)$ satisfies the following DSE
\be
\label{31}
d^{- 1} (\vk) = \frac{1}{g} - I_d (\vk) ,
\ee
where
\be
\label{32}
I_d (\vk) = N_c \int \dbar{q} \bigl[ 1 -  (\hat{\vk} \cdot \hat{\vq})^2 \bigr]
\frac{d(\vk - \vq)}{(\vk - \vq)^2} \frac{1 + 2 n (\vq)}{2 \omega (\vq)} \, .
\ee

As at zero temperature we express the Coulomb propagator
\be
\label{658-X}
F(1, 2) = \langle F_A(1, 2) \rangle
\ee
by means of the Coulomb form factor $f(\vk)$ defined in momentum  space by
\be
\label{663-X}
g^2 F^{ab}(\vk) = g^2 \delta^{ab} \, G(\vk) f(\vk) \, \vk^2 G(\vk) = \delta^{ab} \frac{d(\vk)^2}{\vk^2} \, f(\vk) .
\ee
Using the operator identity
\be
\label{668-X}
F_A = \frac{\partial}{\partial g} (g G_A)
\ee
and, proceeding as in the zero-temperature case (see e.g.\ Ref.~\cite{Feuchter:2004mk}), 
neglecting the $g$-dependence of the density matrix one finds the relation
\be
\label{590-XY}
f(\vk) = - g^2 \frac{\partial}{\partial g} d^{- 1} (\vk)
\ee
from which one finds with Eq.~(\ref{31}) the following integral equation
\begin{align}
\label{75A}
f (\vk) &= 1 + I_f (\vk) \\
\label{1374}
I_f(\vk) &=  N_c \int \dbar{q}  \bigl[ 1 -  (\hat{\vk} \cdot \hat{\vq})^2 \bigr]
\begin{aligned}[t]
&\frac{d(\vk - \vq)^2 \, f(\vk - \vq)}{(\vk - \vq)^2} \\
&\times \frac{1 + 2n(\vq)}{2\omega(\vq)} \, .
\end{aligned}
\end{align}
This equation is formally the same as at zero temperature except that the gluon propagator
is replaced by its finite-temperature counterpart, Eq.~(\ref{488-x2}).


\section{The free energy}\label{section5}

At vanishing chemical potential the thermodynamic potential of the grand canonical
ensemble is given by the free energy
\be\label{524-X1}
\cF = \langle \tilde{H} \rangle - T \cS ,
\ee
where $\cS$ is the entropy, which is defined by
\be\label{17}
\cS = - \kbol \Tr \left[ \frac{\tilde{\cD}}{\tilde{Z}} \: \ln \frac{\tilde{\cD}}{\tilde{Z}} \right] ,
\ee
where
\be\label{18}
\tilde{Z} = \Tr \tilde{\cD}
\ee
is the partition function of the grand canonical ensemble with $\tilde{\cD}$ defined by
Eq.~(\ref{337}). By straightforward algebraic manipulation the entropy Eq.~(\ref{17})
can be converted to 
\be
\label{19}
\cS = \kbol \biggl( \ln\tilde{Z} - \beta \frac{\partial \ln\tilde{Z}}{\partial \beta} \biggr) .
\ee
In principle, we could calculate the free energy from the partition function,
$\tilde{Z} = \exp (- \beta \cF)$. For the exact density matrix Eq.~(\ref{337}) this would
yield the same result as Eq.~(\ref{524-X1}). However, in the present case, where we have
replaced the full density matrix Eq.~(\ref{337}) by the one of a system of independent
quasi-particles, Eq.~(\ref{271}), it is mandatory to evaluate $\cF$ from Eq.~(\ref{524-X1})
in order to capture the essential correlations between the gluons. Besides, the density
matrix $\tilde{\cD}$ Eq.~(\ref{271}) has so far not been determined.

The partition function of the density matrix  $\tilde{\cD}$ [Eq.~(\ref{271})] is
calculated in the standard fashion yielding
\be
\label{23}
\tilde{Z} = \exp \left[ (d-1) (N^2_c-1) V \int \dbar{k} \: \ln\bigl(1 + n(\vk)\bigr) \right] ,
\ee
where $d - 1 = t_{ii}(\vk)$ is the number of independent polarization directions in $d$
spatial dimensions and $N^2_c - 1$ is the number of color degrees of freedom of the
gauge bosons. With Eq.~(\ref{23}) we find from Eq.~\eqref{19} for the entropy density
$s[n]$ per degree of freedom of the gauge bosons, defined by
\be
\label{24}
\cS = (d - 1) (N^2_c - 1) V \frac{1}{(2 \pi)^d} s[n],
\ee
the expression
\be
\label{25}
s[n] = (2 \pi)^d \kbol \int \dbar{k} \left[ \beta n(\vk) \Omega(\vk) + \ln\bigl(1 + n(\vk)\bigr) \right] ,
\ee
where $\Omega(\vk)$ is the variational kernel in the density matrix [Eq.~\eqref{15}].
To obtain the free energy from Eq.~(\ref{524-X1}) we have still to calculate the energy
$\langle \tilde{H} \rangle$, which is done below.

In the evaluation of the expectation value of the Yang--Mills Hamiltonian Eq.~(\ref{398-G1})
we use the same approximation as in the $T = 0$ case keeping only terms up to two loops
in the energy \cite{Feuchter:2004mk}. The magnetic energy is straightforwardly evaluated
using Wick's theorem. One finds
\begin{multline}\label{25-x}
\langle \tilde{H}_B \rangle = \frac{(d-1)(N_c^2-1)}{4} \, V \int \dbar{p} \: \frac{\vp^2}{\omega(\vp)}[1+2n(\vp)] \\
+ \frac{g^2 N_c(N_c^2-1)}{16} \, V \int \dbar{p} \dbar{q} \:
\frac{d(d-3)+3-(\hat{\vp}\cdot\hat{\vq})^2}{\omega(\vp)\omega(\vq)} \\
\times \bigl[1+2n(\vp)\bigr] \bigl[1+2n(\vq)\bigr] .
\end{multline}

The evaluation of the kinetic energy and the Coulomb energy is somewhat more involved. 
With the representation Eq.~(\ref{430}) we obtain from Eq.~(\ref{11-1}) 
\be
\label{16-x}
\tilde{\Pi} (1) = \Pi (1) - \i \, \chi (1, 2) A (2) .
\ee
For the expectation value of the two momentum operators entering the kinetic and Coulomb 
terms one finds by using Eqs.~(\ref{16-x}) and (\ref{488-x2})--(\ref{500-x4})  
\be
\label{22-x}
\langle \tilde{\Pi}^{\dagger} (1) \tilde{\Pi} (2) \rangle = K (1, 2) - \chi (1, 2) + \chi (1, 1') D (1', 2') \chi (2', 2)  .
\ee
Contracting the external indices one obtains from this expression immediately the expectation
value of the kinetic term
\begin{multline}\label{23-x}
\langle \tilde{H}_K \rangle = \frac{1}{2} \langle \tilde{\Pi}^\dagger (1) \tilde{\Pi} (1) \rangle \\
= \frac{1}{2} \left[ K(1,1) - \chi(1,1) + \chi(1,2) D(2,3) \chi(3,1) \right] \\
= \frac{(d-1)(N_c^2-1)}{2} \, V \int \dbar{q} \bigl[ K(\vq) - \chi(\vq) \\ + \chi^2 (\vq) D(\vq) \bigr] .
\end{multline}

This expression is not obviously positive definite (as it should be); however, it is not
difficult to show that this is indeed the case. Using Eqs.~\eqref{488-x2} and \eqref{494-x3}
and separating the zero- and finite-temperature terms, Eq.~\eqref{23-x} can be written as
\begin{multline}\label{23-x-2}
\langle \tilde{H}_K \rangle = \frac{(d-1)(N_c^2-1)}{2} \, V \int \dbar{q}
\Biggl\{ \frac{\bigl[ \omega(\vq)-\chi(\vq) \bigr]^2}{2 \omega(\vq)} \\
+ n(\vq) \: \frac{\omega^2(\vq)+\chi^2(\vq)}{\omega(\vq)} \Biggr\} ,
\end{multline}
from which it is seen that $\langle \tilde{H}_K \rangle$ is indeed positive definite.

Restricting ourselves to including up to two (overlapping) loops in  the energy allows us
to replace the Coulomb kernel $F_A(1, 2)$ in $\tilde{H}_\mathrm{C}$ by its expectation value
Eq.~(\ref{658-X}), resulting in the approximation
\begin{multline}\label{26-x}
\langle \rho(1) F_A(1,2) \rho(2) \rangle \\
\simeq \langle F_A(1,2) \rangle R(1;3,4) R(2;5,6)
\langle A(3) \tilde{\Pi}^\dagger(4) A(5) \tilde{\Pi}(6) \rangle \\
= F(1, 2) R(1;3,4) R(2;5,6) 
\Bigl[
\langle A(3) A(5) \rangle   \langle \tilde{\Pi}^\dagger(4) \tilde{\Pi}(6) \rangle \\
+ \langle A(3)\tilde{\Pi}(6) \rangle \langle \tilde{\Pi}^\dagger(4) A(5) \rangle
\Bigr] .
\end{multline}
The remaining expectation values can be straightforwardly carried out using Eqs.~(\ref{16-x})
and (\ref{22-x}). Then we find for the Coulomb term
\begin{multline}\label{h-495}
\langle\tilde{H}_\mathrm{C}\rangle = \frac{g^2}{2} F(1, 2) R(1;3,4) R(2;5,6) \\
\times \Bigl\{
 D(3,5) \bigl[ K(4,6) - \chi(4,6) + \chi(4,4') D(4',6') \chi(6',6) \bigr] \\
+ \Bigl[ \tfrac{1}{2} \: \delta(3,6) - D(3,3') \chi(3',6) \Bigr] \\
\times \Bigl[ \tfrac{1}{2} \: \delta(4,5) - \chi(4,4') D(4',5) \Bigr]
\Bigr\}.
\end{multline}

As noticed in Ref.~\cite{Reinhardt:2011hq}, with the replacement
\be
\label{324}
F^{ab}_A(\vx,\vy) \to \delta^{ab}
\ee
the Coulomb Hamiltonian $H_\mathrm{C}$ [Eq.~(\ref{398-G1})] becomes
\be
\label{329}
H_\mathrm{C} \to \frac{g^2}{2} J^{- 1}_A Q^a J_A Q^a = \frac{g^2}{2} Q^a Q^a ,
\ee
where
\be
\label{334}
Q^a = \int \d[d]x \: \rho^a(\vx)
\ee
is the total color charge and we have used
\be
\label{339}
[Q^a, J_A] = 0 . 
\ee
The last relation holds since the Faddeev--Popov determinant $J_A$ is
invariant under global color rotations, which are generated by $Q^a$. This is also
explicitly seen by using the representation Eq.~(\ref{430}). In a colorless universe
\be
\label{344-X}
\langle\!\langle  Q^a Q^a \rangle\!\rangle \equiv \langle \tilde{Q}^a \tilde{Q}^a \rangle = \langle {Q}^{a} {Q}^a \rangle = 0
\ee
holds. To ensure that this condition (\ref{344-X}) is respected by $\langle H_\mathrm{C} \rangle$
in Ref.~\cite{Reinhardt:2011hq} the Coulomb kernel $\frac{g^2}{2} F^{ab}_A(\vx,\vy)$ was replaced by
\be
\label{812-X1}
\frac{g^2}{2} F^{ab}_A(\vx,\vy) - \delta^{ab} .
\ee
Furthermore, in Ref.~\cite{Reinhardt:2011hq} the Yang--Mills grand canonical ensemble was
projected onto zero total color. It was found in there that the effects of both color
projection and of the replacement Eq.~(\ref{812-X1}) is negligible. Therefore in the following
we will ignore the projection on zero color states [as well as the replacement Eq.~(\ref{812-X1})].

Rewriting the above given thermal average of the Hamiltonian, Eqs.~(\ref{25-x}), \eqref{23-x}
and \eqref{h-495}, in momentum space one finds for the energy density per degree of freedom
$e [n,\omega]$ defined by
\be\label{33}
\begin{split}
\langle \tilde{H} \rangle &= (d - 1) (N^2_c - 1) \frac{V}{(2 \pi)^d} \: e[n,\omega] , \\
e[n,\omega] &\equiv e_K + e_B + e_\mathrm{C} 
\end{split}
\ee
the following expressions
\begin{widetext}
\begin{subequations}\label{endensity}
\begin{align}
\label{35}
e_K ={}& \frac{(2\pi)^d}{2}
\int \dbar{q} \left[ K(\vq) - \chi(\vq) + \chi^2 (\vq) D(\vq) \right] , \\
\label{36}
e_B ={}& \frac{(2\pi)^d}{2} \, \int \dbar{q} \: \vq^2 \, D(\vq)
+ \frac{g^2 (2\pi)^d N_c (d-1)^2}{4d} \, \int \dbar{p} \dbar{q} \: D(\vp) D(\vq) , \\
e_\mathrm{C} ={}& \frac{g^2 N_c}{4 (d-1)} (2 \pi)^d \int \dbar{p} \dbar{q}
\bigl[ d-2+(\hat{\vp}\cdot\hat{\vq})^2 \bigr] F(\vp + \vq) 
\Bigl\{ 
D(\vp) K(\vq) + D(\vq) K(\vp) 
+ D(\vp) D(\vq) \bigl[ \chi(\vp)-\chi(\vq) \bigr]^2 \nonumber \\
&+ \bigl[ D(\vp)-D(\vq) \bigr] \bigl[ \chi(\vp)-\chi(\vq) \bigr] -\tfrac12 \Bigr\} , \label{37}
\end{align}
\end{subequations}
\end{widetext}
where $F(\vk)$ is the Coulomb propagator defined by Eqs.~\eqref{658-X} and \eqref{663-X}.


\section{\label{section6}The finite-temperature variational principle}

The kernel $\Omega(\vk)$ defining the density matrix $\tilde{\cD}$ [see Eqs.~(\ref{15})
and (\ref{271})] is so far completely arbitrary. We will now determine it by the
finite-temperature variational principle. At fixed temperature and volume and arbitrary
particle number the thermodynamic potential to be minimized is the grand canonical
potential, which in the present case coincides with the free energy since the chemical
potential of the gluons vanishes. Therefore we minimize the free energy or its density
\be
\label{104}
f[n, \omega] = e[n, \omega] - T s[n]  ,
\ee
where $e [n, \omega]$ is the energy density [Eq.~(\ref{33})] and $s[n]$ is the entropy
density [Eq.~\eqref{25}]. Instead of varying the free energy density with respect to
$\Omega(\vk)$ it is more convenient to take the variation with respect to the occupation
numbers $n(\vk)$ [Eq.~(\ref{482-X1})], which is equivalent since $n(\vk)$
is a monotonic function of $\Omega(\vk)$. From Eq.~(\ref{25}) we find for the variation
of the entropy 
\be
\label{106}
\frac{\delta s [n]}{\delta n(\vk)} = \frac{1}{T} \: \Omega(\vk) .
\ee
Therefore stationarity of the free energy density, $\delta f / \delta n = 0$, requires
\be
\label{107}
\Omega(\vk) = \frac{\delta e[n,\omega]}{\delta n(\vk)} \, ,
\ee
which, in the spirit of Landau's Fermi liquid theory, identifies $\Omega(\vk)$ as the
quasi-gluon energy. Of course, this result could have been anticipated from the form of
the finite-temperature Bose occupation numbers, Eq.~(\ref{482-X1}).  

So far the kernel $\omega(\vk)$, which defines the vacuum wave functional [Eq.~(\ref{36-x})]
and thus our basis of the Fock space, is completely arbitrary and we could use any positive
definite kernel and the corresponding gluon basis to calculate the thermodynamic averages.
As long as we include the complete set of states and keep the full canonical density
operator the thermodynamical averages are independent of $\omega(\vk)$. Thus, in principle,
the free energy should not depend on $\omega(\vk)$. However, due to the truncation of the
full Hamiltonian in the density operator $\tilde{\cD}$ [Eq.~(\ref{337})] to a single
particle one [Eq.~(\ref{271})] the actual choice of the basis, i.e.\ of $\omega(\vk)$, does matter and the
optimal choice of $\omega(\vk)$ is obtained by extremizing the free energy
\be
\label{108}
\frac{\delta f[n, \omega]}{\delta\omega(\vk)} = 0 .
\ee
For the evaluation of $\delta e[n,\omega] / \delta\omega(\vk)$ we skip the implicit
$\omega(\vk)$ dependence of $\chi(\vk)$ and $F(\vk)$, since their inclusion would give
rise to higher-order loops. Ignoring this implicit  $\omega(\vk)$ dependence the energy
density depends on $n(\vk)$ and $\omega(\vk)$ only through the gluon field and
momentum propagators, $D$ [Eq.~(\ref{488-x2})] and $K$ [Eq.~(\ref{494-x3})], i.e.
\be
\label{109}
e [n, \omega] = e \left[ D [n, \omega], K [n, \omega] \right] .
\ee
Using the chain rule and the explicit form of the propagators [Eqs.~(\ref{488-x2}) and
(\ref{494-x3})] we obtain
\begin{align}
\label{110}
\frac{\delta e}{\delta n(\vk)} &= \frac{1}{\omega(\vk)} \frac{\delta e}{\delta D(\vk)} +  \omega(\vk) \frac{\delta e}{\delta K (\vk)} , \\
\label{111}
\frac{\delta e}{\delta \omega(\vk)} &= \frac{1 + 2 n(\vk)}{2} \left[ \frac{\delta e}{\delta D(\vk)} - \omega^2(\vk) \frac{\delta e}{\delta K(\vk)} \right] .
\end{align}
Since the entropy density $s[n]$ [Eq.~\eqref{25}] does not explicitly depend on $\omega(\vk)$ the condition Eq.~(\ref{108})
reduces to
\be
\label{112}
\delta e [n, \omega] / \delta \omega(\vk) = 0 \, ,
\ee 
which in view of Eq.~(\ref{111}) leads to the condition
\be
\label{113}
\frac{\delta e}{\delta D(\vk)}  = \omega^2 (\vk) \frac{\delta e}{\delta K(\vk)} \, .
\ee
For $T = 0$, which implies $n(\vk) = 0$, this equation reduces to the gap equation
obtained in Ref.~\cite{Feuchter:2004mk}. 
Inserting the gap equation (\ref{113}) into Eq.~(\ref{110}) we find from Eq.~(\ref{107})
\be
\label{114}
\Omega(\vk) = 2 \omega(\vk) \frac{\delta e}{\delta K(\vk)} \, .
\ee
With the explicit expressions for the energy density given in Eq.~\eqref{endensity} we obtain
\be
\label{115}
\frac{\Omega(\vk)}{\omega(\vk)} = 1 + I_\Omega(\vk) \, ,
\ee
where
\be
\label{116}
I_\Omega(\vk) = \frac{g^2 \, N_c}{2 (d-1)}
\int \dbar{q} \bigl[ d-2+(\hat{\vk}\cdot\hat{\vq})^2 \bigr] F(\vec{k}-\vec{q}) \: \frac{1+2n(\vq)}{{\omega(\vq)}} .
\ee
Using the explicit expressions for the energy density [Eq.~\eqref{endensity}],
the gap equation can finally be expressed as
\be
\label{118}
\omega^2(\vk) = \vk^2 + \chi^2(\vk) + I^{0}_\omega[n] + I_\omega[n] (\vk) ,
\ee
where
\be\label{119-a}
I^{0}_\omega[n]  = \frac{g^2 N_c (d-1)^2}{2d} \int \dbar{q} \: \frac{1+2n(\vq)}{{\omega(\vq)}}
\ee
is the tadpole term stemming from the non-abelian part of the magnetic energy, and
\begin{multline}
\label{119}
I_\omega[n](\vk) = \frac{g^2 N_c}{2 (d - 1)} \int \dbar{q} \bigl[ d - 2 + (\hat{\vk} \cdot \hat{\vq})^2 \bigr] \frac{F(\vk-\vq)}{\omega(\vq)} \\
\times
\Bigl\{
\bigl( 1 + 2 n(\vq) \bigr) \bigl[ \omega^2 (\vq) - \omega^2 (\vk) + \bigl( \chi(\vq) - \chi(\vk) \bigr)^2 \bigr] \\
- 2 \omega(\vq) \bigl( \chi(\vq) - \chi(\vk) \bigr) \Bigr\}
\end{multline}
is the contribution of the Coulomb Hamiltonian.
These loop integrals are ultraviolet (UV) divergent and require regularization and
eventually renormalization of the gap equation. Fortunately the temperature dependence 
of these loop integrals (which is due to the finite-temperature occupation numbers $n(\vk)$)
does not give rise to additional UV singularities. Therefore the zero-temperature
counterterms are, in principle, sufficient to eliminate the UV singularities. However,
special care is required to separate the temperature dependence from the UV-singular terms,
which will be done in the next section.


\section{\label{section7}Renormalization}

At large momenta $\abs{\vk} \gg \kbol T$ the temperature should become irrelevant. Consequently,
the finite-temperature solutions $\omega(\vk)$, $d(\vk)$, $\chi(\vk)$ should possess the
same UV behavior as in the zero-temperature case. Indeed, the finite-temperature
contributions to the loop integrals (the terms proportional to the occupation number
$n(\vk)$) are all ultraviolet finite. This is because for $\abs{\vk} \to \infty$ we have
$\Omega (\vk) \sim \omega (\vk) \sim k$ and thus the finite-temperature occupation numbers
\be
\label{50}
n(\vq\to\infty) \sim \e^{- \beta \Omega(\vq\to\infty)} \sim \e^{- \beta q}
\ee
cut off the large momenta. Therefore, the renormalization can be done independent of the
temperature subtracting only zero-temperature counterterms.

For the renormalization we follow Ref.~\cite{Feuchter:2004mk} and separate the various degrees of
UV divergences of the loop integrals of the gap equation by writing Eq.~(\ref{119}) as
\be
\label{47}
I_\omega[n](\vk) = I_\omega^{(2)}(\vk,T) + 2 \chi(\vk) I_\omega^{(1)}(\vk, T) + \bar{I}_\omega[n](\vk) ,
\ee
where the integrals
\begin{multline}\label{48}
I_\omega^{(l)}(\vk, T) = \frac{g^2 N_c}{2 (d - 1)} \int \dbar{q} \bigl[ d - 2 + ( \hat{\vec{k}} \cdot \hat{\vec{q}} )^2 \bigr]
\frac{F(\vk - \vq)}{\omega(\vq)} \\
\times \Bigl[ \bigl( \omega(\vq) - \chi(\vq) \bigr)^l - \bigl( \omega(\vk) - \chi(\vk) \bigr)^l \Bigr]
\end{multline}
are linearly (for $l = 1$) and quadratically (for $l = 2$) UV divergent, while the
finite-temperature contribution 
\begin{multline}\label{49}
\bar{I}_\omega[n](\vk) = \frac{g^2 N_c}{d - 1} \int \dbar{q} \bigl[ d - 2 + (\hat{\vec{k}} \cdot \hat{\vec{q}})^2 \bigr]
\frac{F(\vk - \vq)}{\omega(\vq)} \\
\times n(\vq) \Bigl[ \omega^2(\vq) - \omega^2(\vk) + \bigl(\chi(\vq) - \chi(\vk) \bigr)^2 \Bigr]
\end{multline}
is UV convergent due to Eq.~(\ref{50}). Analogously we write for the tadpole Eq.~(\ref{119-a})
\be\label{531*}
I_\omega^0 [n] = I_\omega^0 + \bar{I}_\omega^0[n], \qquad I_\omega^0 \equiv I_\omega^0[n=0] .
\ee
The loop integrals $I_\omega^{(l)}(\vk,T)$ and $I_\omega^0$ are defined as in the
zero-temperature case, Ref.~\cite{Feuchter:2004mk}. However, their entries $\omega(\vk)$, $d(\vk)$,
and $\chi(\vk)$ are temperature dependent. These integrals contain all UV divergences,
while the finite-temperature modifications $\bar{I}_\omega[n](\vk)$ and $\bar{I}_\omega^0 [n]$
are UV finite. Therefore, the renormalization can be carried out, in principle, in the
same way as in the zero-temperature case, Ref.~\cite{Reinhardt:2007wh}. However, due to the implicit
temperature dependence of $\omega(\vk), d(\vk), \chi(\vk)$ we have to be careful not to
introduce finite-temperature effects by the renormalization. Following the renormalization
procedure given in Ref.~\cite{Reinhardt:2007wh} for $T = 0$ and subtracting only zero-temperature terms
one arrives at the renormalized gap equation
\begin{multline}\label{52}
\omega^2(\vk) = \vk^2 + \overline{\chi}^2(\vk) + \Delta I_\omega^{(2)}(\vk) + \Delta I_\omega ^{0}  + \bar{I}_\omega^{0}[n] + c_0 \\
+ 2 \overline{\chi}(\vk) \bigl[ \Delta I_\omega ^{(1)}(\vk) + c_1 \bigr] + \bar{I}_\omega[n](\vk) ,
\end{multline}
where
\begin{align}
\label{53}
\overline{\chi}(\vk)  &= \chi (\vk) -  \chi (\mu_\chi) |_{T = 0} \nonumber\\
\Delta I_\omega^{(l)} (\vk) &= I_{\omega}^{(l)} (\vk) - I_{\omega\rvert_{T = 0}}^{(l)} (0) \nonumber\\
\Delta I_\omega^{0}  &= I_\omega^0 - I_{\omega\rvert_{T = 0}}^0
\end{align}
and $c_0$, $c_1$ are finite renormalization constants surviving from energy counterterms
\cite{Epple:2007ut}
\be\label{cntterms}
\Delta H = \frac12 \, C_0 \, A(1) A(1) + \mathrm{i} C_1 A(1) \Pi(1) .
\ee
$c_0$ and $c_1$ are the finite parts of the divergent constants $C_0$ and $C_1$, respectively,
i.e.\ $C_i = C_i^\mathit{div}+c_i$.
Furthermore, $\mu_\chi$ is an arbitrary scale arising from the
renormalization of the Faddeev--Popov determinant \cite{Epple:2007ut}. The subscript $T = 0$ in Eq.~\eqref{53}
means that the corresponding quantities have to be taken with the zero-temperature
self-consistent solution. The choice of these finite renormalization parameters will be
discussed in Sect.~\ref{sec:choice}.


The renormalized ghost DSE is obtained from Eq.~(\ref{31}) with a subtraction at an
arbitrary scale $\mu_d$ at $T=0$
\be\label{31-REN}
\begin{split}
d^{-1} (\vk) &= d^{-1}(\mu_d) - \Delta I_d (\vk), \\
\Delta I_d (\vk) &= I_d(\vk) - I_d(\mu_d)\big\rvert_{T = 0} \, .
\end{split}
\ee
The Gribov-Zwanziger confinement scenario requires the horizon condition 
\be
\label{1131-1}
d^{- 1} (k = 0)\big\rvert_{T = 0} = 0 ,
\ee
to be satisfied at $T = 0$. This condition can be explicitly built in the renormalized
ghost DSE~(\ref{31-REN}) by choosing the renormalization constant $d (\mu_d)$ such that
\be
\label{1066}
d^{- 1}(\mu_d) = \Delta I_d (k = 0)\big\rvert_{T = 0} = I_d (0)\big\rvert_{T = 0} - I_d (\mu_d)\big\rvert_{T = 0} \, ,
\ee
which can be fulfilled for arbitrary renormalization scale $\mu_d$. Inserting this value
into Eq.~(\ref{31-REN}) we obtain
\be
\label{1073-Z}
d^{- 1}(\vk) = - \bigl( I_d (\vk) - I_d (0) \big\rvert_{T = 0} \bigr) ,
\ee
which is nothing but the renormalized ghost DSE~(\ref{31-REN}) with the renormalization
scale fixed at $\mu_d = 0$ [and the horizon condition Eq.~\eqref{1131-1} built in]. This
shows that implementing the horizon condition automatically puts the renormalization
scale in Eq.~(\ref{31-REN}) to $\mu_d = 0$. Solutions of the coupled DSEs and gap equation satisfying the horizon 
condition (\ref{1131-1}) are called ``critical'' and those with $d^{- 1} (0) > 0$ subcritical \cite{Epple:2007ut}.

The renormalized DSE for the Coulomb form factor $f (\vk)$ [Eq.~(\ref{663-X})] is analogously
obtained by subtracting Eq.~(\ref{75A}) once at $T = 0$ and an arbitrary renormalization
scale $\mu_f$, yielding
\be\label{75A-REN}
\begin{split}
f (\vk) &= f (\mu_f)  +\Delta I_f (\vk), \\
\Delta I_f (\vk) &= I_f(\vk) - I_f(\mu_f) \bigr\rvert_{T = 0} \,.
\end{split}
\ee
In principle, the renormalization scale $\mu_f$ of the Coulomb form factor
can be independently chosen from that of the ghost $\mu_d$. However, since $f(\vk)$ and
$d(\vk)$ are tightly related by Eq.~(\ref{590-XY}) for consistency one should choose $\mu_f = \mu_d$. 

Within our approach the finite-temperature Yang--Mills theory is now determined by the
following set of coupled equations: The Eq.~(\ref{115}) for the quasi-gluon energy
$\Omega(\vk)$, the gap equation (\ref{52}) for $\omega(\vk)$, the DSE~(\ref{31-REN})
for the ghost form factor $d(\vk)$ and the DSE (\ref{75A-REN}) for the Coulomb
form factor $f(\vk)$. 

A comment is here in order: In principle, we would get only two equations from the
finite-temperature variational principle, one equation for $\Omega(\vk)$ and one for
$\omega(\vk)$. However, in the process of evaluating $\langle \tilde{H} \rangle$ we have
introduced, for convenience, additional propagators (ghost and Coulomb propagator), which
we did not explicitly express as functionals of $\omega(\vk)$ (and $\Omega(\vk)$).
Instead of that we derived DSEs for these quantities, which all contain loop integrals.
In taking the variation of the energy $\langle \tilde{H} \rangle$ with respect to $\omega(\vk)$
and $\Omega(\vk)$ the implicit $\omega (\vk)$\,---\,and $\Omega (\vk)$\,---\,dependence
of these propagators is ignored since it would give rise to two-loop terms in the
equations of motion, see Ref.~\cite {Campagnari:2010wc} for more details.


\section{Infrared analysis}\label{section8}

Before presenting the numerical solutions of the coupled Eqs.~(\ref{115}), (\ref{52}),
\eqref{1073-Z} and \eqref{75A-REN}, we investigate their infrared behavior. At zero
temperature the infrared analysis was carried out in Refs.~\cite{Feuchter:2004mk,Schleifenbaum:2006bq}. At
arbitrary finite temperature the infrared analysis of the DSEs cannot be done in the
usual way. This is because the finite-temperature occupation numbers $n(\vk)$
[Eq.~(\ref{482-X1})] depend exponentially on the gluon quasi-energy $\Omega(\vk)$. The
infrared analysis can, however, be carried out in the high-temperature limit where the
occupation numbers become
\be
\label{58}
n(\vk) \overset{T\to\infty}{\sim} \frac{\kbol T}{\Omega(\vk)} \, .
\ee
This limit is sufficient to exhibit the infrared behavior of the propagators in the
deconfined phase. With the representation Eq.~(\ref{58}) the infrared analysis of the
coupled Dyson--Schwinger equations can essentially be carried out as at $T = 0$.

In general, the IR analysis can be carried out in two ways: i) using the angular
approximation \cite{Feuchter:2004mk} replacing kernels $K(\vk - \vq)$ depending on the difference
between the external momentum $\vk$ and the loop momentum $\vq$ by
\be
\label{657}
K (\vk - \vq) \to K(k) \, \Theta(k - q) + K(q) \, \Theta(q - k) ,
\ee
with $k=\abs{\vk}$, $q=\abs{\vq}$.
Although being approximative, this method has the advantage that IR power law ans\"atze
for the propagators are indeed restricted to the IR regime. ii) Without using the angular
approximation the IR power law ans\"atze have to be used in the loop integrals for the whole
momentum range \cite{Lerche:2002ep,Schleifenbaum:2006bq}. This method is, in principle, exact in the IR (up
to the omission of possible logarithms) but has the disadvantage that the UV behavior of
these integrals is usually changed by the IR power law ans\"atze and UV-finite loop
integrals may turn into UV-divergent ones. Fortunately both methods yield very similar
results as we will explicitly demonstrate in the Appendix. 

For the infrared analysis we assume the following power law ans\"atze
\be
\label{1145-3}
\omega(\vk\to0) \sim \frac{A}{k^\alpha} \, , \qquad d(\vk\to0) \sim \frac{B}{k^\beta} .
\ee
For $T = 0$ the infrared analysis with and without angular approximation was carried out
in Refs.~\cite{Feuchter:2004mk} and \cite{Schleifenbaum:2006bq}, respectively. Assuming the horizon condition
[Eq.~(\ref{1131-1})] and a bare ghost-gluon vertex one finds from the ghost DSE~(\ref{31}) 
in $d$ spatial dimensions in both cases the sum rule
\be
\label{1136-2}
\alpha = 2 \beta + 2 - d  .
\ee
Including also the gap equation the infrared exponents are fixed and one finds,
abandoning the angular approximation, in $d = 3$ spatial dimensions two solutions \cite{Schleifenbaum:2006bq}
\be
\label{1155-4}
\beta \simeq 0.8, \qquad \beta = 1 .
\ee
(In the angular approximation only the second solution is obtained.) Both exponents are
also found in the numerical solutions, and were originally obtained in Ref.~\cite{Feuchter:2004mk}
($\beta\simeq0.8$) and in Ref.~\cite{Epple:2006hv} ($\beta = 1$). In $d = 2$ spatial
dimensions one finds a single solution with $\beta = 1/2$ in the angular approximation
and $\beta = 0.4$ when the angular approximation is abandoned, while the numerical
solution \cite{Feuchter:2007mq} yields $\beta \simeq 0.44$. 

For high temperatures using the approximation Eq.~(\ref{58}) the infrared analysis is
carried out in the Appendix. From the ghost DSE one finds the same sum rule Eq.~(\ref{1136-2})
as in the zero-temperature case. Including also the gap equation one obtains in $d = 3$
at high temperatures a solution with the infrared exponent of the ghost form factor
\be
\label{1170-5}
\beta = 1/2 \, 
\ee
both with and without the angular approximation. With this value for $\beta$ one finds
from the infrared analysis of the DSE for the Coulomb form factor $f(\vk\to0) \sim 1/k^\lambda$
the infrared exponent $\lambda \approx 1$ which together with $\beta = 1/2$ for the
ghost form factor leads to a linearly rising Coulomb potential $F(\vk\to0) \sim 1/k^4$,
as is also found in the lattice calculation \cite{Nakagawa:2006fk}. The infrared
exponents obtained in the infrared analysis are confirmed by the numerical calculations,
as we will see in Section~\ref{section10}.


\section{Neglect of the Coulomb term}\label{section9}

\begin{figure*}[t]
\centering
\includegraphics[width=.8\linewidth]{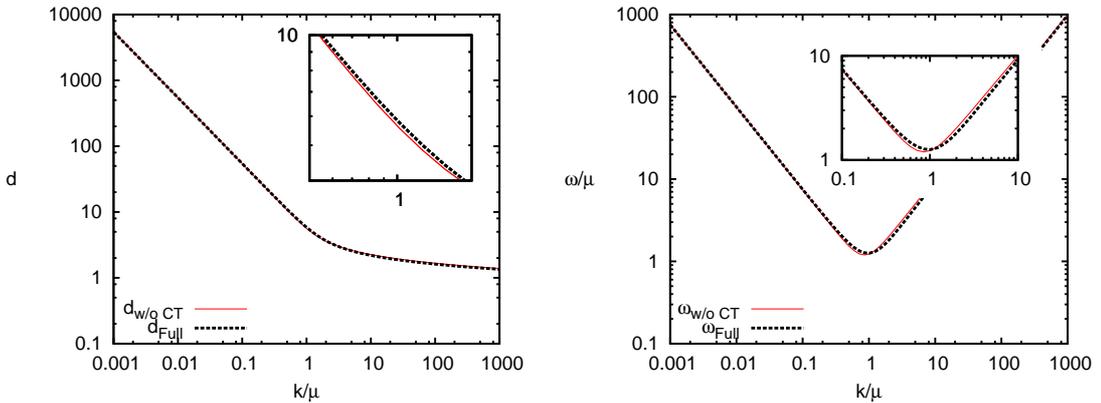}
\caption{The ghost form factor (left panel) and the gluon energy (right panel) at $T=0$
with and without the Coulomb term.}
\label{fig:compareCT}
\end{figure*}

\begin{figure}[t]
\centering
\includegraphics[width=.7\linewidth]{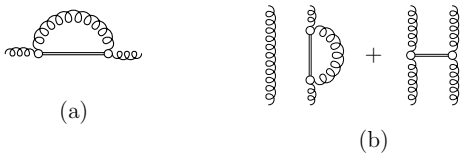}
\caption{\label{fig:CouDiag}Contributions of the Coulomb term to (a) the gluon self-energy and (b) the
two-gluon energy. A wavy line represents the gluon propagator, the double line stands
for the Coulomb kernel.}
\end{figure}

We are mainly interested in the description of the deconfinement phase transition. If the
Gribov confinement scenario is realized in Coulomb gauge, this transition should manifest
itself in a change of the infrared behavior of the various propagators, in particular in
that of the ghost form factor $d(\vk)$ and the gluon energy $\omega(\vk)$ and $\Omega(\vk)$,
respectively. In the (renormalized) gap equation (\ref{52}) the contributions from the
Coulomb term are infrared sub-leading ($\Delta I^{(l)} (k = 0) = 0$) even for infrared
finite $\omega(\vk)$. Furthermore, the ghost DSE~(\ref{31}) does not receive contributions
from the Coulomb term. Therefore, for a first qualitative description of the deconfinement
phase transition the Coulomb term should be negligible. With the neglect of the Coulomb
term, the gap equation (\ref{52}), reduces to
\be
\label{55}
\omega^2 (\vk) = \vk^2 + \overline{\chi}^2 (\vk) + \Delta I_\omega^{0} + \bar{I}^{0}_\omega [n] + c_0 + 2 c_1 \overline{\chi}(\vk) ,
\ee
while the ghost DSE~\eqref{1073-Z} remains unchanged. At $T = 0$ both $\Delta I^0_\omega$ [Eq.~(\ref{53})]
and $\bar{I}^0_\omega [n]$ [Eq.~(\ref{531*})] vanish and the gap equation (\ref{55})
reduces to
\be
\label{56}
\omega^2 (\vk) = \vk^2 + \overline{\chi}^2 (\vk) + 2 c_1 \overline{\chi}(\vk) + c_0\, .
\ee
Figure~\ref{fig:compareCT} shows the result of the numerical solution of the ghost DSE
and the full gap equation\footnote{When the Coulomb term is included the DSE for the
Coulomb form factor has also to be solved. This was done as described in Ref.~\cite{Feuchter:2004mk}
replacing $d(\vk)$ by its bare value $d(\vk) = 1$ in the loop integral $I_f(\vk)$.}%
~(\ref{52}) at $T = 0$ and the corresponding solutions of the gap equation (\ref{56})
with the Coulomb term neglected for $c_0=c_1=0$.
As is seen, there are only small deviations in the
mid-momentum regime, while both solutions agree in the ultraviolet and, in particular,
in the infrared. Furthermore, in a quasi-particle description of the gluon sector
(underlying the present approach) the neglect of the Coulomb term is conceptually  
advantageous (and expected to give a better description) as long as two particle
correlations are neglected. The reason is the following: The Coulomb term gives rise to
the UV-singular quasi-gluon self-energy, $\Delta \Omega(\vk) = \omega (\vk) I_\Omega (\vk)$,
see Eq.~(\ref{115}), which is diagrammatically illustrated in Fig.~\ref{fig:CouDiag}a.
In a two-(quasi-)gluon state these divergent self-energy contributions are precisely
canceled by the divergent contribution from the Coulomb interaction to the two-gluon
energy, shown in Fig.~\ref{fig:CouDiag}b.
Therefore it does not make sense to keep the
Coulomb term in the quasi-gluon energy as long as the two-body correlations are not
taken into account.\footnote{Note also that in QED the Coulomb term vanishes identically
in the absence of external charges.} We will therefore neglect the Coulomb term in the
following. Then the loop integral in Eq.~(\ref{115}) has to be discarded so that
\be
\label{56-1}
\Omega (\vk) = \omega (\vk) .
\ee
Furthermore, $\Delta I^{0}_\omega$ [Eq.~(\ref{53})] represents the change of the tadpole
due to the change of the self-consistent solution $\omega (\vk)$ at finite temperature
(relative to the zero-temperature case) but does not contain the change of the tadpole
due to the explicit temperature dependence of the gluon propagator via the
finite-temperature occupation numbers. Therefore, we expect $\Delta I^{0}_\omega$ to be
small and we will neglect it. Then the finite-temperature gap equation (\ref{55}) reduces to 
\be
\label{57}
\omega^2 (\vk) = \vk^2 + \overline{\chi}^2 (\vk) + \bar{I}_\omega^{0} [n] + 2 c_1 \, \overline{\chi}(\vk) + c_0 \, .
\ee 
Here only $\bar{I}^0_\omega[n]$ (the temperature-dependent part of the tadpole, see Eq.~(\ref{531*}))
is explicitly temperature dependent, while the curvature $\chi(\vk)$ depends only
implicitly on the temperature via the ghost form factor. However, as we will see in
Sect.~\ref{section10}, it is this temperature dependence of the ghost which triggers the
deconfinement phase transition.


\section{\label{sec:choice}Choice of renormalization constants}

Let us now discuss the choice of the finite renormalization parameters.
Since the renormalization can be completely accomplished by renormalizing the theory at
zero temperature, the remaining finite-temperature renormalization constants $c_0$, $c_1$
should be also fixed at $T=0$ and then kept fixed independent of the temperature. To fix
these constants we notice that the $T=0$ solutions are IR divergent with
$\omega(k) \sim \chi(k) \sim 1/k^\alpha$, $\alpha=0.8$, $1$, see Sect.~\ref{section8}.
For such solutions the gap equation \eqref{56} reduces in the IR with Eq.~\eqref{1145-3} to
\be\label{drc1}
\omega(k) - \overline{\chi}(k) = c_1 + \frac{c_0}{2A} \: k^\alpha + \dots
\ee
The constant $c_1$ obviously determines the infrared limit of $\omega(\vk) - \overline{\chi}(\vk)$, 
and $c_1 = 0$ is required in order that the 't~Hooft loop obeys an area law, Ref.~\cite{Reinhardt:2007wh}.
This value is also favored by the variational principle (of minimal energy), Ref.~\cite{Reinhardt:2007wh}.
Furthermore, recent lattice investigations of the Yang--Mills vacuum wave functional in
$d = 2$ spatial dimensions show that for this value of $c_1$ the wave functional
Eq.~(\ref{36-x}) yields statistical weights for abelian plane-wave configurations which
are in very good agreement with the ones of the exact vacuum wave functional \cite{Greensite:2011pj}. 
We will therefore put $c_1 = 0$.

As is seen from Eq.~\eqref{drc1} the renormalization constant $c_0$ is IR subleading
compared to $c_1$ and influences the mid-momentum regime, which is however expected to
affect the deconfinement phase transition. Fortunately in the present variational approach
$c_0$ needs not to be treated as a free parameter but can be determined by minimizing the
energy density. First notice that with $c_1=0$ the IR behavior of the $T=0$ gap equation
\eqref{drc1} reduces to
\be\label{drc2}
\omega(k) - \overline{\chi}(k) = \frac{c_0}{2A} \: k^\alpha + \dots
\ee
Furthermore, with Eq.~\eqref{430}, which is correct to the two-loop order (in the energy)
considered in the present paper, our vacuum wave functional defined by Eqs.~\eqref{8-x}
and \eqref{36-x} reads
\be\label{drc4}
\langle A | 0 \rangle = \mathcal{N} \exp \left\{ - \frac{1}{2} \, A(1) \bigl[ \omega(1,2) - \overline{\chi}(1,2) \bigr] A(2) \right\} .
\ee
In order that this wave functional is regular $\lvert\langle A | 0 \rangle\rvert^2<\infty$
for all $A$ we must have
\be\label{Z1}
\omega(k) - \overline{\chi}(k) \geq 0
\ee
According to Eq.~\eqref{drc2} this requires for small $k$
\be\label{Z2}
c_0 \geq 0 .
\ee
We will see that this condition is also required by the positivity
of the energy. Furthermore the numerical results given in Sect.~\ref{section10} will show
that the energy density takes its minimal value for $c_0=0$.

Using the gap equation \eqref{56} the zero-temperature energy density Eq.~\eqref{endensity}
with the counterterm $\sim C_0$ [Eq.~\eqref{cntterms}] fully included\footnote{The counterterm
$\sim C_1$ [Eq.~\eqref{cntterms}] is only needed for the Coulomb energy.} can be cast into the form
\be\label{Z3}
e = e_K + e_B = \frac{(2\pi)^d}{2} \int \dbar{k} \bigl[ \omega(\vk) - \overline{\chi}(\vk) \bigr] .
\ee
Note that the energy density does not explicitly but merely implicitly depend on $c_0$,
since $\omega(\vk)$ and $\overline{\chi}(\vk)$ depend on $c_0$ via the gap equation \eqref{56}.
From Eq.~\eqref{Z3} it is seen that stability of our non-perturbative vacuum requires
again the condition \eqref{Z1}, i.e.\ $c_0\geq0$.

\begin{figure}
\includegraphics[width=.8\linewidth]{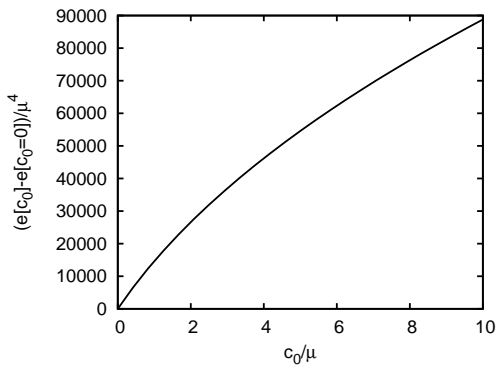}
\caption{\label{fig:deceC0func} Energy density [Eq.~\eqref{Z3}] as function of the
renormalization constant $c_0$ for the solution with IR exponent $\alpha=1$.}
\end{figure}

Figure~\ref{fig:deceC0func} shows the energy density $e$ [Eq.~\eqref{Z3}] as function of $c_0$
for the solution with IR exponent $\alpha=1$.
The energy density Eq.~\eqref{Z3} takes its minimal value at $c_0=0$, which is therefore
the physical value and thus chosen in alla variational calculations.

There is one hidden 
renormalization constant left, which is the scale $\mu_\chi$ in the subtracted curvature
$\overline{\chi} (\vk)$ Eq.~(\ref{53}). This quantity can be chosen in a wide range without
significantly changing the resulting propagators, see Ref.~\cite{Feuchter:2004mk}. (Clearly, this
parameter cannot be chosen $\mu_\chi = 0$ since this would result in $\omega (0) = 0$,
which is in contradiction to the Gribov scenario and also with the findings on the
lattice \cite{Burgio:2009xp}.)


\section{Numerical results}\label{section10}

The coupled ghost DSE~(\ref{1073-Z}) and the gap equation (\ref{57}) can be solved
numerically for the whole momentum range by iteration in the way described in
Ref.~\cite{Epple:2006hv} for the zero-temperature case. (For details see also Ref.~\cite{Leder:2010ji}.)
In the following we will restrict ourselves to $d=3$ spatial dimensions and to $N_c=2$.
For the numerical calculation it is convenient to introduce dimensionless quantities,
rescaling all dimensionful quantities with appropriate powers of an arbitrary momentum
scale $\mu$. The rescaled dimensionless quantities will be indicated by a bar:
\be\label{1325}
\overline{k} \coloneq \frac{k}{\mu} \,, \qquad \overline{\omega}(\overline{k}) \coloneq \frac{\omega(\overline{k} \mu )}{\mu}\,,\qquad \text{etc.}
\ee
Because of the use of a logarithmic momentum grid an infrared cut-off $\lambda_{\mathrm{IR}}$
is needed, which we choose, in dimenionless units, in the range $\overline{\lambda}_{\mathrm{IR}} = 10^{-8} \ldots 10^{-5}$.


\subsection{Zero-temperature solutions}\label{section10a}

Before presenting the finite-temperature solutions, it is worthwhile and necessary to
reconsider the zero-temperature case, from which the physical scale is fixed. We keep the remaining undetermined renormalization
constant fixed at $\bar{\mu}_\chi = 4$, but its precise value is irrelevant
for the qualitative behavior of the obtained solutions.\footnote{Note that, in principle,
the value of this finite renormalization constant could also be determined by minimizing
the free energy with respect to this parameter. This is, however, numerically extremely
expensive and beyond the scope of the present paper.}

\begin{figure*}[t]
\centering
\includegraphics[width=0.7\linewidth]{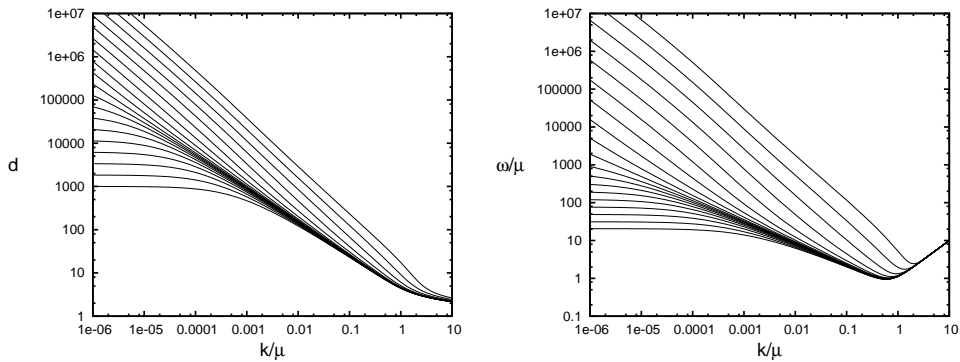}
\caption{The ghost form factor (left panel) and the gluon energy (right panel) for
different values of $d^{-1}(\lambda_{\mathrm{IR}})$ at $T=0$}.
\label{fig:variousD0}
\end{figure*}

The Gribov--Zwanziger confinement scenario requires $d^{-1}(k=0)=0$. Solutions which
satisfy this condition are referred to as ``critical'', while solutions with
$d^{-1}(k=0)>0$ are called ``subcritical''. Figure~\ref{fig:variousD0} shows the results
of the numerical solutions of the ghost DSE~(\ref{31-REN}) and the gap equation (\ref{57})
at $T = 0$  and various choices of the ghost renormalization constant $d^{- 1} (\mu_d = \lambda_\mathrm{IR})$, 
starting at $d^{-1}(\lambda_{\mathrm{IR}})> 3 \overline{\lambda}_{\mathrm{IR}}$ and successively decreasing
$d^{-1}(\lambda_{\mathrm{IR}})$. As long as $d^{-1}(\lambda_{\mathrm{IR}}) \gtrsim 2 \overline{\lambda}_{\mathrm{IR}}$ the solutions 
obtained are subcritical, i.e.\ $d^{- 1}(0) > 0$. At sufficiently small
$d^{-1}(\lambda_{\mathrm{IR}}) \approx 2 \overline{\lambda}_{\mathrm{IR}}$ the critical solution with
$\beta\simeq0.8$ is obtained. When $d^{-1}(\lambda_{\mathrm{IR}})$ is further decreased
critical solutions with $\beta =1$ appears.

The critical solutions can be more easily found by implementing the horizon condition
$d^{-1}(k=0)=0$ explicitly into the ghost DSE~(\ref{31-REN}) by choosing its
renormalization constant $d^{- 1} (\mu_d = 0) = 0$ [resulting in Eq.~(\ref{1073-Z})]
and solving the coupled equations (\ref{1073-Z}) and (\ref{57}) iteratively. Choosing
as starting point of the iteration $d^{-1}(\lambda_{\mathrm{IR}}) \gtrsim \overline{\lambda}_{\mathrm{IR}}$
one obtains the critical solution with $\beta\simeq0.8$, while the choice $d^{- 1} (\lambda_{\mathrm{IR}}) \lesssim \overline{\lambda}_{\mathrm{IR}}$
results in a continuous set of critical solutions with $\beta=1$ differing in the value
of the IR coefficient $B$ of $d(\vk)$, see Eq.~(\ref{1145-3}). 
Let us also mention that the convergence of the iteration is generally faster when the
initial functions $d(\vk)$ and $\omega(\vk)$ satisfy the IR sum rule Eq.~(\ref{1136-2})
and the relation (\ref{1971-G41A}), which represents the IR limit of the ghost DSE, see
the Appendix. For $\beta = 1$ and $d = 3$ Eq.~(\ref{1971-G41A}) reduces to
\be
\frac{A}{B^2} = \frac{N_c}{ 8 \pi^2}\,.
\label{amplitsIR}
\ee

\begin{figure*}[t]
\centering
\includegraphics[width=.7\linewidth]{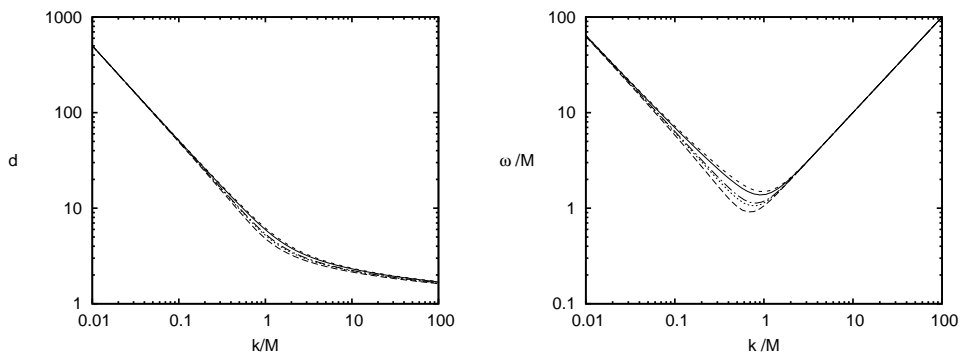}
\caption{The different critical solutions with $\beta =1$ overlayed to the same infrared
coefficient $B$.}
\label{fig:differentB}
\end{figure*}

In Fig.~\ref{fig:differentB} the infrared coefficients $B$ [Eq.~(\ref{1145-3})] of the
various $\beta=1$ solutions have been rescaled to make them coincide in the IR, so they
differ in the mid-momentum regime.

As already mentioned before our equations can be entirely expressed in terms of
dimensionless quantities, reflecting the scale invariance of Yang--Mills theory. To
confront our numerical results with the experimental or lattice data we have to fix the
(so far arbitrary) scale $\mu$, see Eq.~(\ref{1325}). For this purpose we use the
lattice data for the gluon propagator, which were obtained by fixing the scale by means
of the Wilsonian string tension $\sigma_\mathrm{W} \simeq (440\,\mbox{MeV})^2$.
(This quantity has not yet been calculated in the present Hamiltonian approach.)
The lattice calculations carried out in Coulomb gauge in Ref.~\cite{Burgio:2009xp} show
that the gluon energy $\omega(\vk)$ can be nicely fitted by Gribov's formula \cite{Gribov:1977wm}
\be\label{eq:GribForm}
\omega(\vk) = \sqrt{\vk^2 + \frac{M^4}{\vk^2}},
\ee
with $M \simeq 880\, \text{MeV} \simeq 2 \sqrt{\smash[b]{\sigma_\mathrm{W}}}$.
The Gribov formula implies the IR exponent $\alpha =1$ [cf.~Eq.~(\ref{1145-3})] and
hence by the IR sum rule (\ref{1136-2}) $\beta =1$. Therefore, to fix our scale from
the lattice result for $M$, we have to use the $\beta =1$ solutions. As discussed above,
there is a whole set of $\beta = 1$ critical solutions differing in the IR coefficient
of the ghost form factor $\bar{B} = B/\mu$. From these we chose the one which fits
Gribov's formula Eq.~(\ref{eq:GribForm}) best. For the above adopted values $\overline{\mu}_\chi = 4$
and $c_0 =0$ one finds for this solution the value
\be
\bar{B} = \frac{B}{\mu} = 6.01 \pm 0.05 .
\ee
From the Gribov formula Eq.~(\ref{eq:GribForm}) one reads off the IR coefficient
(\ref{1145-3}) of the gluon energy 
\be\label{1430-XX}
A=M^2 .
\ee
With this result we find from Eq.~(\ref{amplitsIR}) for our scale
\be
\mu = \frac{M}{\bar{B}} \sqrt{\frac{8 \pi^2}{N_c}} .
\label{eqs:scaledelta}
\ee
This fixes our physical scale in terms of the Gribov mass $M$.


\subsection{Finite-temperature solutions}\label{section10b}

In this section we use the same procedure described above to solve the coupled equations
(\ref{1073-Z}) and (\ref{57}) for finite temperatures. The renormalization described in
Sect.~\ref{section7} requires subtractions with the zero-temperature solutions in order
to have temperature independent renormalization constants. However, such subtractions are
numerically unstable. Because of this, in the numerical results presented in this section
the subtraction of the ghost loop has been done with the finite-temperature solutions.
Consequently, the horizon condition $d^{-1}(k=0)=0$ is built in explicitly at any temperature. 

\begin{figure}[t]
\centering
\includegraphics[width=.8\linewidth]{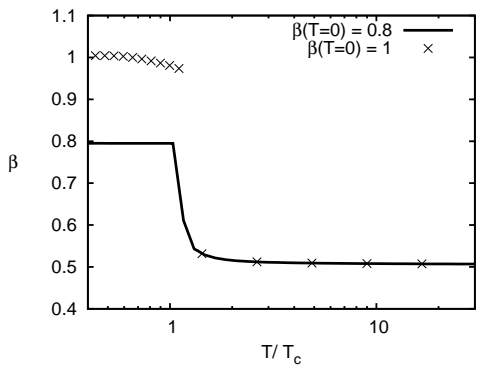}
\caption{Infrared exponent of the ghost form factor as function of the temperature for
the critical solutions at $T = 0$ with $\beta\simeq0.8$ (full line) and $\beta=1$ (crosses).
Above $T_c$ both solutions merge to a single solution.}
\label{fig:betafunction}
\end{figure}

Starting with the critical solutions at $T = 0$ and increasing the temperature the
self-consistent solutions remain more or less unchanged up to a critical temperature $T_c$,
where the IR exponent of the ghost form factor suddenly changes to $\beta \approx 0.5$
and the gluon energy becomes infrared finite $(\alpha \approx 0)$ in accord with the sum rule Eq.~(\ref{1136-2}). 
This is nicely seen in Fig.~\ref{fig:betafunction} where we show the infrared exponent
for the critical solutions as function of the temperature.
The two critical solutions with $\beta\simeq0.8$ and $\beta = 1$ existing below $T_c$
merge to a single solution which approaches $\beta = 0.5$ for $T \gg T_c$. The IR exponent
$\beta = 0.5$ is precisely the value of the ghost form factor in $d = 2$. Thus, the change
of the IR exponent at $T_c$ is in accord with dimensional reduction.

\begin{figure}[t]
\includegraphics[width=.8\linewidth]{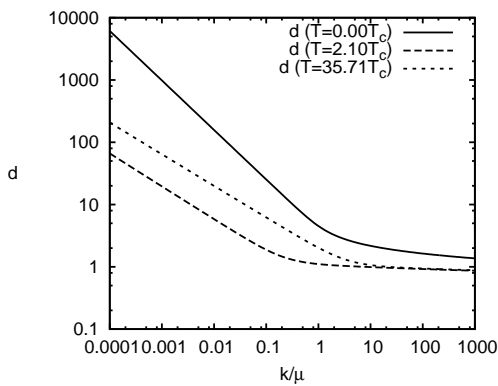}
\caption{Solutions of the ghost DSE for different temperatures $T$.}
\label{fig:finiteTGhost}
\end{figure}

\begin{figure}
\includegraphics[width=.8\linewidth]{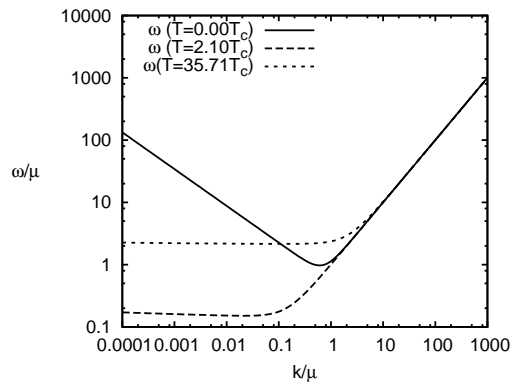}
\caption{Solutions of the gluon gap equation for different temperatures $T$.}
\label{fig:finiteTGluon}
\end{figure}

\begin{figure}
\includegraphics[width=.8\linewidth]{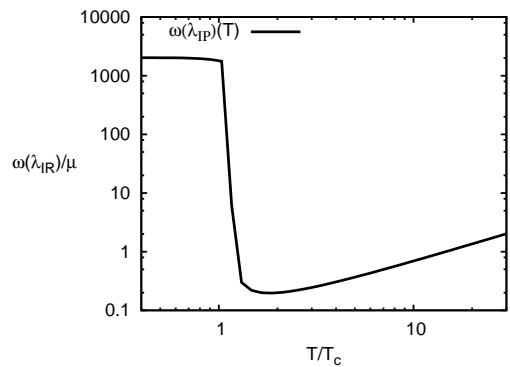}
\caption{ $\omega (\lambda_{\mathrm{IR}})$ as function of the temperature.}
\label{fig:infraredGluon}
\end{figure}

Figures~\ref{fig:finiteTGhost} and \ref{fig:finiteTGluon} show the self-consistent
finite-temperature solutions for the ghost form factor and the gluon energy as function
of the momentum for various temperatures.
As the temperature is increased above $T_c$ the plateau value of the gluon energy starts
increasing linearly with the temperature as is seen in Fig.~\ref{fig:infraredGluon},
where we show the infrared value $\omega (\lambda_{\mathrm{IR}})$ as function of the
temperature. The linear increase with the temperature
of $\omega (\lambda_{\mathrm{IR}})$ is easily understood by noticing that above $T_c$ the temperature is the only energy
scale and for dimensional reason $\omega (\lambda_{\mathrm{IR}})$ has therefore to scale with the temperature.

\begin{figure}
\includegraphics[width=.8\linewidth]{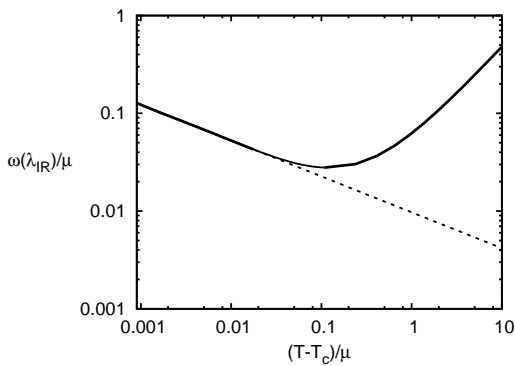}
\caption{Critical behavior of the effective gluon mass for $T \searrow T_c$.}
\label{fig:criticalGmass}
\end{figure}

The infrared value of the gluon energy $\omega (\lambda_{\mathrm{IR}})$ can be interpreted as an effective gluon mass. 
Zooming into the behavior of $\omega (\lambda_{\mathrm{IR}})$ near $T_c$, which is done in Fig.~\ref{fig:criticalGmass}, we can extract the
critical exponent $\kappa$ of this quantity defined by 
\be
\label{1558-8}
\frac{\omega(\lambda_{\mathrm{IR}})}{\mu} = \lk \frac{T}{T_c} - 1 \rk^{- \kappa} .
\ee
From our numerical solution we extract a value of $\kappa \approx 0.37$, which is similar
to the value of $\kappa\approx0.41$ obtained in Ref.~\cite{Castorina:2011ja}, where a
quasi-gluon picture has been used to fit the lattice results for the energy density and
the pressure, and furthermore using critical exponents from the $d = 3$ Ising model, which is in the same 
universality class as SU$(2)$ gauge theory.

The sudden change of the ghost infrared exponent, see Fig.~\ref{fig:betafunction}
(or alternatively the drop in the IR value of the gluon energy $\omega (\lambda_{\mathrm{IR}})$, see Fig.~\ref{fig:infraredGluon}) 
can be used to determine the critical temperature $\bar{T}_c$. From our numerical solutions we extract with
$T_c =\mu \bar{T}_c$ and Eq.~(\ref{eqs:scaledelta}) for SU(2) the value of 
\be
\label{1547-7}
T_c \sim 275 \cdots 290 \, \text{MeV} \, , 
\ee
which is somewhat smaller than the lattice result of $T_c \simeq 295 \, \mbox{MeV}$ [SU(2)]. 

\begin{figure}
\includegraphics[width=.8\linewidth]{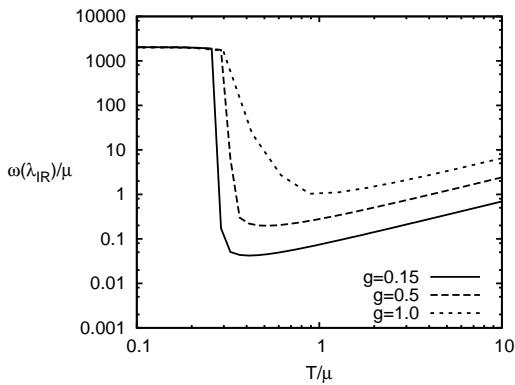}
\caption{ $\omega (\lambda_{\mathrm{IR}})$ as function of the temperature for various
values of the coupling $g$.}
\label{fig:varygc2}
\end{figure}

The value of the critical temperature is seen to be insensitive of the actual value of
$0< d^{-1}(\mu_d=\lambda_{\mathrm{IR}}) \approx \overline{\lambda}_{\mathrm{IR}}$.
Yet there is one parameter which can effect the form of the phase transition: the
coupling constant $g$ entering the tadpole term $\bar{I}_\omega^0[n]$.
This term acts as an effective mass in the gap equation \eqref{57}. As
long as the temperature is small, we expect the IR behavior
of the gluon energy to be insensitive of the precise value of this effective mass. However, when the temperature
is raised the effect of the mass scale cannot be neglected.
When we increase the value of $g$ (Fig.~\ref{fig:varygc2}) the phase transition is weakened.
For the self-consistent solutions we have chosen $g=0.5$.

Removing the tadpole, the free QED solution ($\omega(\vk) = k$, $\alpha =-1$ and $\beta= 0$) exists at
high temperatures.

\begin{figure}[t]
\includegraphics[width=.8\linewidth]{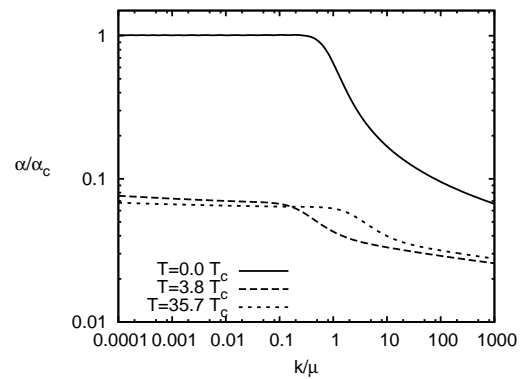}
\caption{Zero- and finite-temperature running coupling constant normalized to the infrared value $\alpha_c$ at $T=0$.}
\label{fig:runningCfinite}
\end{figure}

In Fig.~\ref{fig:runningCfinite} we show the running coupling constant $\alpha (k)$ 
calculated from the ghost-gluon vertex, as described in Ref.~\cite{Schleifenbaum:2006bq}, for both
$T = 0$ and $T > T_c$ and normalized to the infrared value $\alpha_c$ (at $T=0$). In both
cases $\alpha (k)$ is IR finite but the IR plateau value decreases by an order of magnitude
above the deconfinement phase transition. Below and above $T_c$ this plateau value
remains more or less unchanged as the temperature varies. The IR finiteness of the
running coupling is guaranteed by the sum rule Eq.~\eqref{1136-2} of the IR exponents of
ghost and gluon propagators, which holds both at $T=0$ and $T\to\infty$. As is well known,
in the high-temperature limit $d=4$ Yang--Mills theory reduces to $d=3$ Yang--Mills theory
coupled to a Higgs field, which is the temporal component of the gauge field in $d=4$. Since the
Higgs field certainly contributes to the running coupling constant there is no obvious
reason why at high temperatures the running coupling should approach that of the (confining)
$d=3$ Yang--Mills theory, although it is IR finite in both $d=4$ Yang--Mills theory
at $T\to\infty$ and $d=3$ Yang--Mills theory at $T=0$.

\begin{figure}
\includegraphics[width=.8\linewidth]{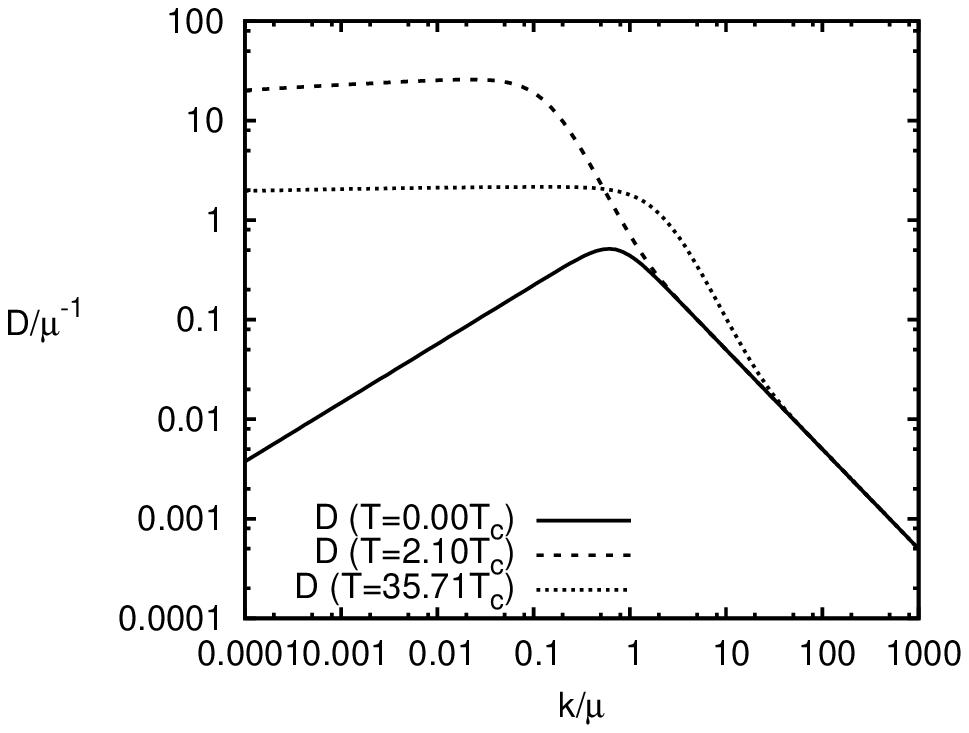}
\caption{Gluon propagator $D(\vk)$ [\Eqref{488-x2}] at zero and finite temperature.}
\label{fig:GluonProp}
\end{figure}

Finally, in Fig.~\ref{fig:GluonProp} we show the gluon propagator \Eqref{488-x2}
at zero and at finite temperatures above the deconfinement phase transition.
Below the deconfinement transition temperature the gluon propagator is in accord with
Gribov's formula \eqref{eq:GribForm} vanishing in the IR, while above the deconfinement
phase transition the gluon is massive. Furthermore the effective IR gluon masses increase
with the temperature as is explicitly seen in Fig.~\ref{fig:varygc2}. Note also that in
the UV the gluon propagator is independent of the temperature.


\section{Summary and Conclusions}\label{section11}

In this paper we have studied the grand canonical ensemble of Yang--Mills theory in the
Hamiltonian approach in Coulomb gauge and investigated the finite-temperature deconfinement
phase transition. For the density operator a quasi-particle picture was assumed and the
quasi-particle energies were determined by minimizing the free energy. A complete basis
of the gluonic Fock space was constructed by creating an arbitrary number of quasi-gluons
on top of a Gaussian-type vacuum wave functional whose width was considered as variational
kernel and determined by minimizing the free energy. This results in the finite-temperature
gap equation, which has to be solved together with the Dyson--Schwinger equations for the
ghost and Coulomb propagators. We have shown that the effect of the Coulomb term of the
Yang--Mills Hamiltonian (which represents the longitudinal part of the kinetic energy)
is negligible. Neglecting this term the gap equation and DSE for the ghost propagator
decouple from the DSE for the Coulomb propagator. We have solved these equations
analytically in the high-temperature limit in the infrared regime. We have found that
the infrared exponents of the gluon energy ($\alpha$) and of
the ghost form factor ($\beta$) satisfy in the high-temperature limit the same sum rule
$\alpha = 2 \beta - 1$ as at zero temperature. While at zero temperature two solutions with $\beta = 1$ and
$\beta\simeq0.8$ exist, in the high-temperature limit there is only a single solution $\beta = 1/2$. Our numerical
solution of the coupled ghost DSE and gap equation shows that at a critical temperature $T_c$ there is a sudden change
of the infrared exponents from their zero-temperature values to their high-temperature limits. At this deconfinement
phase transition both $T = 0$ critical solutions with $\beta\simeq0.8$ and $\beta = 1$ merge to a single solution with 
$\beta = 1/2$. In accord with the sum rule at the critical temperature the gluon energy changes from being
infrared divergent ($\alpha\simeq0.6$ and $\alpha = 1$) to being infrared finite ($\alpha = 0$). This shows that while the gluons are absent
from the infrared spectrum in the confined phase they become massive particles in the deconfined phase. For the effective
gluon mass we have found a critical exponent of $0.37$. From the sudden change of the infrared exponents and of the
infrared value of the gluon mass we have extracted a critical temperature of the deconfinement phase transition of 
$T_c = 275 \dots 290$~MeV using the lattice results for the Gribov mass.

An alternative way to describe the deconfinement phase transition is to study the behavior
of the Polyakov loop, as was done in the FRG approach in Landau gauge \cite{Braun:2007bx}.
This can be also done in the present approach and will be subject to future work.

Towards the extension of the present approach to full QCD one has first to study the
gauge group $SU(3)$, which has a first order phase transition. From the lattice studies
carried out in Ref.~\cite{Maas:2011ez} in Landau gauge one may expect that the order of
the phase transition manifests itself in the critical behavior of the effective gluon
mass close to the phase transition, but this is an open issue and requires further studies.

The results obtained in the present paper are rather encouraging for an extension of the
present approach to full QCD at finite temperature and baryon density. A first step in
this direction was recently undertaken by studying chiral symmetry breaking in a
variational approach to QCD in Coulomb gauge at zero temperature and density \cite{Pak:2011wu}.
The extension of this approach to finite temperature and non-vanishing chemical potential
is, in principle, straightforward but technically involved. Since this approach is
formulated in the continuum it will not face the problems one encounters in the lattice
formulation at finite chemical potential.


\begin{acknowledgments}
The authors would like to thank P.~Watson for a critical reading of the manuscript and
A.~S.~Szczepaniak for discussions on ongoing projects. This work was supported by the
Deutsche Forschungsgemeinschaft (DFG) under Contracts Nos Re856/6-3 and Re856/9-1 and by
the BMBF under Contract No.\ PT-GSI-06TU7199.
\end{acknowledgments}


\appendix*

\section{Infrared analysis}

Below we carry out the IR analysis of the coupled DSEs in the high-temperature limit. 
As usual we assume power law ans\"atze in the infrared ($k=\abs{\vk}$)
\begin{align}
\label{h-1067-2}
\omega (k) &= \frac{A}{k^\alpha} \, ,&
d (k) &= \frac{B}{k^\beta} \, , &
\chi (k) &= \frac{C}{k^\gamma} \, , \nonumber \\
\Omega (k) &= \frac{E}{k^\varepsilon} \, , &
f (k) &= \frac{L}{k^\lambda}
\end{align}
and, in addition, for the ghost form factor $d(k)$ the horizon condition
\be
\label{ha-1145}
d^{- 1} (k = 0) = 0 ,
\ee
so that $\beta > 0$.   The coefficients $A$, $B$\ldots as well as the IR exponents $\alpha$,
$\beta$\ldots are expected to depend on the temperature. The modifications at finite
temperature arise exclusively from the extra piece of the gluon propagator Eq.~(\ref{488-x2})
containing the finite-temperature occupation numbers $n(k)$ Eq.~(\ref{482-X1}), defined
in terms of the kernel $\Omega(k)$ occurring in the ansatz for the density matrix,
Eq.~\eqref{15}. Obviously, for $\varepsilon > 0$ the finite-temperature part of the
gluon propagator is IR subleading since in this case
\be
\label{1676-X}
n(k\to0) = 0
\ee
and we expect the zero-temperature result for the sum rules of the IR exponent to remain
true at finite $T$. Note also that the property (\ref{1676-X}) is maintained in the
high-temperature approximation Eq.~(\ref{58}). 

\subsection{Infrared analysis in angular approximation}

We begin with the analysis of the Dyson--Schwinger equation for the ghost form factor.
It differs from the zero-temperature equation only by the replacement of the zero-temperature
gluon propagator $1/[2 \omega(k)]$ by its finite-temperature counterpart $[1 + 2n(k)]/[2\omega(k)]$.
As in the $T = 0$ case it is convenient to consider the derivative of the ghost DSE \eqref{31}
\be
\label{h-1078-4}
\frac{\d{}}{\d k} d^{- 1} (k) = I'_d (k) .
\ee
In the angular approximation, Eq.~(\ref{657}), after the angular integration over the $d-1$-sphere
\be
\label{1273x}
\frac{1}{{(2\pi)^d }} \int\limits_{S_ {d - 1}} \d^{d - 1} \Omega
\bigl[ 1 - (\hat{\vec{k}}\cdot \hat{\vec{q}})^2 \bigr] = \frac{d-1}{(4\pi)^{d/2}}\frac{1}{\Gamma(1+\frac{d}{2})}
\ee
we obtain for the loop integral (\ref{32}) 
\be
\label{h-1083-5}
I'_d (k) = \frac{N_c}{2} \frac{d-1}{(4\pi)^{d/2}}\frac{1}{\Gamma(1+\frac{d}{2})} \left(  \frac{d (k)}{k^2} \right)' R (k) ,
\ee
where
\be
\label{h-1088-6}
R (k) =  \int^k_0 \d q \,q^{d-1} D (q) = \int^k_0 \d q\, q^{d-1}\frac{  1 + 2 \frac{\kbol T}{\Omega (q)} }{\omega (q)}
\ee
and we have used here the high-temperature limit Eq.~(\ref{58}). For small $k$ we can
safely use the infrared asymptotic forms (\ref{h-1067-2}) for $\omega(k)$ and $\Omega(k)$
in the integrand, which yields
\be
\label{h-1094-7}
R (k) = \frac{k^{\alpha + d}}{A (\alpha + d)} \left[ 1 + \frac{\kbol T}{E} \: k^\varepsilon \frac{\alpha + d}{\alpha + \varepsilon + d} \right] .
\ee
Inserting this result into (\ref{h-1083-5}) we find from (\ref{h-1078-4}) the relation
\begin{multline}
\label{h-1099-8}
\frac{A}{B^2} = \frac{N_c }{2}  \frac{d-1}{(4\pi)^{d/2}}\frac{1}{\Gamma(1+\frac{d}{2})} \frac{\beta + 2}{\beta (\alpha + d)} 
\: k^{\alpha - 2 \beta + d - 2} \\
\times \left[ 1 + \frac{\kbol T}{E} \: k^\varepsilon \frac{\alpha + d}{\alpha + \varepsilon + d} \right] ,
\end{multline}
where the left-hand side is a constant for fixed $T$. For $\varepsilon \geq 0$ the second term in the bracket is not
IR leading compared to the first one and we obtain the relation
\be
\label{h-1109-9}
\alpha = 2 \beta - d + 2 \, ,
\ee 
which is just the zero-temperature infrared sum rule \cite{Schleifenbaum:2006bq}. In the opposite case,  $\varepsilon < 0$ the second
term is IR leading and the sum rule
\be
\label{h-1109-10}
\alpha + \varepsilon = 2 \beta - d + 2
\ee
follows. Note the two sum rules merge continuously at $\varepsilon = 0$. 

Consider now Eq.~(\ref{115}) for the gluon quasi-energy $\Omega(k)$. Obviously, 
if the Coulomb term is neglected, we have $\Omega(k)=\omega(k)$ and consequently $\varepsilon = \alpha$.
For the infrared analysis of the full equation it is again convenient to take the derivative 
\be
 \label{X2-5}
 \frac{\d{}}{\d k} \left( \frac{\Omega (k)}{\omega (k)} \right) = I^\prime_\Omega (k) .
 \ee
With the ans\"atze (\ref{h-1067-2}) the left-hand side yields
\be
 \label{X4-15}
 \frac{\d[]}{\d k} \left( \frac{\Omega (k)}{\omega (k)} \right) = \frac{D}{A} \left( \alpha - \varepsilon \right) k^{\alpha - \varepsilon - 1} ,
\ee
Using the angular approximation (\ref{657}) for the Coulomb kernel $F ( \vk - \vq)$ we find for the derivative of the loop integral
\be
\label{h-1123-12}
I'_\Omega (k) = \frac{g^2 N_C}{2(4\pi)^{d/2}}\frac{d-1}{\Gamma(1+\frac{d}{2})}  F' (k) R (k) \, ,
\ee
where $R (k)$ has been defined by Eq.~(\ref{h-1088-6}). Using the explicit form of the Coulomb kernel Eq.~(\ref{663-X})
\be
\label{h-1212}
g^2 F (k) = \frac{d (k)^2 f(k)}{k^2}
\ee
we obtain with the ans\"atze (\ref{h-1067-2}) the infrared behavior
\be
\label{h-1128-14}
F (k) = \frac{B^2 L}{k^{2 \beta + 2 + \lambda}} \, .
\ee
With this expression and (\ref{h-1094-7}) we find from Eq.~(\ref{h-1123-12}) 
\begin{multline}
\label{dec-760}
I'_\Omega (k) = - \frac{g^2 N_c}{2(4\pi)^{d/2}}\frac{d-1}{\Gamma(1+\frac{d}{2})} \frac{B^2 L}{A} \: k^{\alpha - 2 \beta + d - 3 -\lambda} \\
\times \frac{2 \beta + 2+\lambda}{\alpha + d} \left[1 + \frac{\kbol T}{E} \frac{\alpha + d}{\alpha + \varepsilon + d} \: k^\varepsilon \right] .
\end{multline}
Inserting Eqs.~(\ref{X4-15}) and (\ref{dec-760}) into (\ref{X2-5}) 
we find for $\varepsilon \geq0$ the relation
\be
  \varepsilon = 2\beta -d + 2 +\lambda\,,\ee
while for $\varepsilon <0$
\be
  2\varepsilon = 2\beta -d + 2 +\lambda\,
\ee
holds.
Using the results (\ref{h-1109-9}) and (\ref{h-1109-10}) from the ghost equation we find in both cases the sum rule
\be
\label{1778}
   \varepsilon = \alpha + \lambda\,.
\ee

In the case of an IR finite Coulomb form factor ($\lambda =0$) the latter equations implies
\be
  \varepsilon = \alpha\, ,
\ee
so that $\Omega(k)$ and $\omega(k)$ have the same IR behavior, up to possible logs.
Indeed with $\lambda = 0$ and $\varepsilon = \alpha > 0$ satisfying the sum rule
(\ref{h-1109-9}) we find from Eq.~(\ref{dec-760}) 
\be
\label{dec-766}
I'_\Omega (k) \sim k^{- 1},
\ee
implying that $\Omega(k)$ has an extra infrared logarithm in addition to the infrared
power law inherited from $\omega(k)$ \footnote{Note that the Coulomb integral $I_\Omega (k)$
[Eq.~(\ref{116})] is positive definite since its integrand is positive, so that $\Omega(k)$,
like $\omega(k)$, is positive definite.}
\be
\label{dec-771}
\Omega (k) = \omega (k) (1 + \sim \ln k) , \qquad k \to 0 
\, .
\ee
The expression (\ref{30}) for the curvature $\chi (k)$ is formally the same as in the
zero-temperature case and consequently its infrared analysis can be carried out as in
the zero-temperature case, Ref.~\cite{Feuchter:2004mk}. In $d$ spatial dimensions one
obtains from the derivative of Eq.~(\ref{30}) for the curvature with the ans\"atze (\ref{h-1067-2}) 
\be
\label{1807}
\frac{C}{B^2} = \frac{N_c}{2 \, (4 \pi)^{d/2}} \frac{1}{\Gamma (1 + d/2)} \frac{\beta + 2}{\gamma
 (d - \beta)} \: k^{d - 2 \beta - 2 + \gamma} ,
\ee
from which we find
\be
\label{h-1118-11}
\gamma = 2 \beta - d +2 \, ,
\ee
which is the zero-temperature result. 
Inserting this relation into eqs. (\ref{h-1109-9}) and (\ref{h-1109-10}) we obtain
\be\label{1819}
\begin{split}
\alpha & = \gamma , \qquad \varepsilon \geq 0 \\
\alpha + \varepsilon & = \gamma , \qquad \varepsilon < 0 \, .
\end{split}
\ee
Consider now the finite-temperature gap equation (\ref{52}). Proceeding as in 
the zero-temperature case \cite{Feuchter:2004mk} one easily shows that for IR divergent $\chi(k)$, i.e.\ $\gamma > 0$, the gap equation 
(\ref{h-1067-2}) reduces in the IR limit to
\be
\label{1832}
\omega^2 (k \to 0) = \chi^2 (k \to 0)
\ee
implying that in the ans\"atze (\ref{h-1067-2}) we have
\be
\label{1837}
A = C .
\ee
With this equality the left-hand sides of Eq.~(\ref{h-1099-8}) (ghost DSE) and of Eq.~(\ref{1807})
(curvature) become equal. Equating the right-hand side of these equations and using the sum rules
(\ref{h-1109-9}) and (\ref{h-1118-11}) to cancel the IR power-laws we obtain the relation
\be
\label{1835}
\frac{1}{d - 1} \frac{\beta + 2}{\alpha ( d - \beta)} = \frac{\beta + 2}{\beta (\alpha + 2)} \, .
\ee
Using the sum rule (\ref{h-1109-9}) again to eliminate $\alpha$ in favor of $\beta$
we finally obtain
\be
\label{1844}
1 = \frac{2}{d - 1} \frac{\beta (\beta + 1)}{(2 \beta + 2 - d) (d - \beta)} \, .
\ee
The solutions of this equation are
\be\label{1849-XX}
\begin{aligned}
\beta & = \frac{1}{2}  & (d &= 2), \\
\beta & = 1  & (d &= 3) ,
\end{aligned}
\ee
which were quoted in Sect.~\ref{section8}. If the curvature $\chi (k)$ is IR finite,
i.e.\ $\gamma \leq 0$, the IR dominant terms on the right-hand side of the gap equation (\ref{52})
is non-zero due to the temperature-dependent part of the tadpole, $\bar{I}^0 [n]$.  
Therefore in this case $\omega(k)$ is IR finite and non-vanishing, implying $\alpha = 0$.
With this value the sum rule (\ref{h-1109-9})  yields
\be
\label{1858}
\beta = \frac{d - 2}{2} \, ,
\ee
i.e.
\be
\label{1863}
\beta = \frac{1}{2} \quad (d = 3) .
\ee
This is the solution found in the numerical calculations in the deconfined regime, while
the solutions (\ref{1849-XX}) are realized in the confined phase. 


\subsection{Infrared power law ans\"atze in loop integrals}
 
Using the power law ans\"atze (\ref{h-1067-2}) for
the whole momentum regime the loop integrals (\ref{32}), (\ref{48}) and (\ref{55}) can be calculated
analytically. For the zero-temperature case this has been done in Refs.~\cite{Schleifenbaum:2006bq} and \cite{Epple:2007ut}. 
With the high-temperature limit of the occupation numbers Eq.~(\ref{58}) these results can be extended to finite temperatures. 
In this subsection, for simplicity,
 we will neglect the Coulomb term entirely, so that $\Omega(k)=\omega(k)$. 

Using the infrared ans\"atze (\ref{h-1067-2}) one finds from the ghost DSE (\ref{31})
\begin{multline}\label{eqGhostIRhighT}
\frac{A}{B^2}=-\frac{N_c(d-1) }{4}\frac{1}{(4\pi)^{d/2} }  k^{d -2 +\alpha - 2\beta  } \\
\times\left[ I_{d \omega}(\alpha,\beta,d) +k^\alpha \frac{2 \kbol T}{A} I_{d \omega}(2\alpha,\beta,d)\right] ,
\end{multline}
where \cite{Schleifenbaum:2006bq}
\be
\label{1958}
  I_{d \omega}(\alpha,\beta,d)=\frac{\Gamma(\frac{d}{2} + \frac{\alpha}{2}) \Gamma(\frac{d}{2} - \frac{\beta}{2}) \Gamma(
  \frac{1}{2}(2 - d - \alpha + \beta))}{
\Gamma(1 - \frac{\alpha}{2}) \Gamma(d + \frac{\alpha}{2} - \frac{\beta}{2}) \Gamma(
  1 + \frac{\beta}{2})}\,.
\ee
The new (finite-temperature) element is the second term in the bracket of Eq.~(\ref{eqGhostIRhighT}). 
At $T = 0$ or for $\alpha > 0$ (and arbitrary $T$) this term is IR subleading and one finds the sum rule
\be
\label{1881-X}
\alpha = 2 \beta - d + 2  \, 
\ee
already previously obtained in angular approximation, see Eq.~(\ref{h-1109-9}). 
With this sum rule (and $\alpha > 0$) Eq.~(\ref{eqGhostIRhighT}) reduces in the IR to
\be
\label{1971-G41A}
\frac{A}{B^2} = - \frac{N_c (d - 1)}{4} \frac{1}{(4 \pi)^{d/2}} \: I_{d \omega} (\alpha, \beta, d) \, .
\ee
Putting here $\beta = 1$ and $d = 3$ we obtain
\be
\label{1892}
\frac{A}{B^2} = \frac{N_c}{8 \pi^2} \, ,
\ee
which is the relation (\ref{amplitsIR}) quoted in Sect.~\ref{section10a}.

For $T > 0$ and $\alpha < 0$ the second term in Eq.~(\ref{1849-XX}) becomes IR dominant and 
we find the sum rule
\be
\label{1981-G41}
2 \alpha = 2 \beta - d + 2 \, ,
\ee
which was also obtained in the previous subsection in the angular approximation.

We now turn to the gap equation (\ref{55}). The neglect of the Coulomb terms is irrelevant for its IR analysis since
these terms are IR subleading as discussed in the previous subsection.

Using the power law ansatz (\ref{h-1067-2}) for the ghost form factor in the whole momentum
regime one finds for the ghost loop Eq.~(\ref{30})
\be
\chi (k)  = k^{d -2 - 2\beta} {B^2}  \frac{N_c}{4}\frac{1}{(4\pi)^{d/2} } I_{d d}(\beta,d) ,
\label{eqGluonIRhighT}
\ee
where
\be
I_{d d}(\beta,d) =
\frac{\Gamma(\frac{d}{2} - \frac{\beta}{2})^2 \Gamma(\beta+1-\frac{d}{2})}%
     {\Gamma(1 + \frac{\beta}{2})^2  \Gamma(d -\beta)}\,.
\ee
From (\ref{eqGluonIRhighT}) we read off the IR exponent Eq.~(\ref{h-1067-2}) of $\chi (k)$ to be given by
\be
\label{2013-G42}
\gamma = 2 \beta - d + 2 \, ,
\ee
which is again the same result as obtained in the angular approximation (\ref{h-1118-11}). We can distinguish now two cases:
\begin{itemize}
\item[i)] If the ghost loop $\chi (k)$ is IR divergent, i.e.\ $\gamma > 0$, the gap equation (\ref{55}) reduces in the IR to
\be
\label{2020-G42A}
\omega^2 (k) = \chi^2 (k)
\ee
and we find $\gamma = \alpha$, which also follows by combining (\ref{1881-X}) and (\ref{2013-G42}) and agrees again with the results obtained in the angular
approximation.

Since $\omega (k)$, $\chi (k) > 0$, Eq.~(\ref{2020-G42A}) reduces then to
\be
\label{2030}
\omega (k) = \chi (k) .
\ee
Inserting here for $\chi (k)$ the expression (\ref{eqGluonIRhighT}) and using the sum rule (\ref{1881-X}) one finds the relation
\be
\label{2035}
\frac{A}{B^2} = \frac{N_c (d - 2)}{8} \frac{1}{(4 \pi)^{d / 2}} I_{dd} (\beta, d) .
\ee
Combining this equation with Eq.~(\ref{1971-G41A}) and using in the latter the sum rule (\ref{1881-X}) we obtain the condition
\be
 I_{d d}(\beta,d)= -\frac{1}{2} I_{d \omega} (\alpha, \beta, d) \Big\rvert_{\alpha = 2 \beta - d + 2} \,.
\label{eqGhostGluonImplicit}
\ee
In $d = 3$ spatial dimensions this equation has the two solutions \cite{Schleifenbaum:2006bq}
\be
\label{2045}
\beta = 1 , \qquad \beta \simeq 0.795 .
\ee
Only the first solution is found in angular approximation, see Eq.~(\ref{1849-XX}).
Both solutions are found numerically in the confined phase, see Refs.~\cite{Feuchter:2004mk}
and \cite{Epple:2006hv}, respectively. 
\item[ii)] If $\chi (k)$ is IR finite or vanishing the finite-temperature part of the
tadpole $\bar{I}^0_\omega [n]$ in the gap equation (\ref{55}) guarantees that $\omega (k)$ is
IR finite but not vanishing, i.e.\ $\alpha = 0$. In this case we find from the sum rule
(\ref{1881-X})
\be
\label{2054}
\beta = \frac{d - 2}{2} \, .
\ee
For $d = 3$ this yields $\beta = \frac{1}{2}$, which is the solution realized in the deconfined phase, as our numerical solutions show.
\end{itemize}


\end{document}